\documentclass[sigconf]{acmart} % ,review,anonymous
\AtBeginDocument{%
  }

\usepackage{xspace}
\usepackage{subcaption}
\usepackage{enumitem}

\copyrightyear{2025}
\acmYear{2025}
\setcopyright{cc}
\setcctype{by}
\acmConference[UIST '25]{The 38th Annual ACM Symposium on User Interface Software and Technology}{September 28-October 1, 2025}{Busan, Republic of Korea}
\acmBooktitle{The 38th Annual ACM Symposium on User Interface Software and Technology (UIST '25), September 28-October 1, 2025, Busan, Republic of Korea}\acmDOI{10.1145/3746059.3747600}
\acmISBN{979-8-4007-2037-6/2025/09}

\newcommand{\name}
{\textsc{DxHF}\xspace}
\acmJournal{JDS}
\acmVolume{37}
\acmNumber{4}
\acmArticle{111}
\acmMonth{8}

\usepackage{ifthen}
\newboolean{revising}
\setboolean{revising}{false}
\ifthenelse{\boolean{revising}}
{
    \newcommand{\rv}[1]{\textcolor{blue}{#1}}
} {
    \newcommand{\rv}[1]{#1}
}

\acmSubmissionID{6644}

\begin{document}

% \title{Interactive Aligned Decomposition for LLM Alignment}
% \title{Decomposition Can Make More Accurate \\Human Feedback for LLM Alignment}
% \title{\name: Improving Accuracy of Human Feedback for LLM Alignment with Interactive Decomposition}
\title[DxHF: Providing High-Quality Human Feedback for LLM Alignment]{DxHF: Providing High-Quality Human Feedback for \\LLM Alignment via Interactive Decomposition}


\author{Danqing Shi}
\affiliation{%
 \institution{Aalto University}
 \city{Helsinki}
 \country{Finland}}
 \email{shidanqingnet@gmail.com}

\author{Furui Cheng}
\affiliation{%
  \institution{ETH Zürich}
  \city{Zürich}
  \country{Switzerland}}
  \email{furui.cheng@inf.ethz.ch}

\author{Tino Weinkauf}
\affiliation{%
  \institution{KTH Royal Institute of Technology}
  \city{Stockholm}
  \country{Sweden}}
  \email{weinkauf@kth.se}

\author{Antti Oulasvirta}
\affiliation{%
  \institution{Aalto University}
 \city{Helsinki}
 \country{Finland}}
  \email{antti.oulasvirta@aalto.fi}

\author{Mennatallah El-Assady}
\affiliation{%
  \institution{ETH Zürich}
  \city{Zürich}
  \country{Switzerland}}
  \email{menna.elassady@ai.ethz.ch}

\renewcommand{\shortauthors}{Shi et al.}
%%
%% Article type: Research, Review, Discussion, Invited or position
\acmArticleType{Review}
%%
%% Links to code and data
\acmCodeLink{https://github.com/borisveytsman/acmart}
\acmDataLink{htps://zenodo.org/link}
%%
%% Authors' contribution
\acmContributions{BT and GKMT designed the study; LT, VB, and AP
  conducted the experiments, BR, HC, CP and JS analyzed the results,
  JPK developed analytical predictions, all authors participated in
  writing the manuscript.}
%%
%% Sometimes the addresses are too long to fit on the page.  In this
%% case uncomment the lines below and fill them accodingly.
%%
%% \authorsaddresses{Corresponding author: Ben Trovato,
%% \href{mailto:trovato@corporation.com}{trovato@corporation.com};
%% Institute for Clarity in Documentation, P.O. Box 1212, Dublin,
%% Ohio, USA, 43017-6221}

\keywords{AI Alignment, User Interface, Human-AI Interaction, Large Language Models, Reinforcement Learning from Human Feedback}

\begin{abstract}
Human preferences are widely used to align large language models (LLMs) through methods such as reinforcement learning from human feedback (RLHF). 
However, the current user interfaces require annotators to compare text paragraphs, which is cognitively challenging when the texts are long or unfamiliar.
This paper contributes by studying the decomposition principle as an approach to improving \rv{the quality of human feedback for LLM alignment.}
This approach breaks down the text into individual claims instead of directly comparing two long-form text responses. 
Based on the principle, we build a novel user interface \name.
It enhances the comparison process by showing decomposed claims, visually encoding the relevance of claims to the conversation and linking similar claims. 
This allows users to skim through key information and identify differences for better and quicker judgment.
Our technical evaluation shows evidence that decomposition generally improves feedback accuracy regarding the ground truth, particularly for users with uncertainty.
A crowdsourcing study with 160 participants indicates that using \name improves feedback accuracy by an average of 5\%, although it increases the average feedback time by 18 seconds. Notably, accuracy is significantly higher in situations where users have less certainty.
The finding of the study highlights the potential of HCI as an effective method for improving human-AI alignment. 
\end{abstract}

\ccsdesc[500]{Human-centered computing~Human computer interaction (HCI)}
% \ccsdesc[500]{Human-centered computing~Visualization}

\maketitle

% Author Keywords
\keywords{Human-AI interaction; LLM alignment; Human feedback}

\section{Introduction}

\begin{figure}[!t]%
%\centering%
\includegraphics[width=\linewidth]{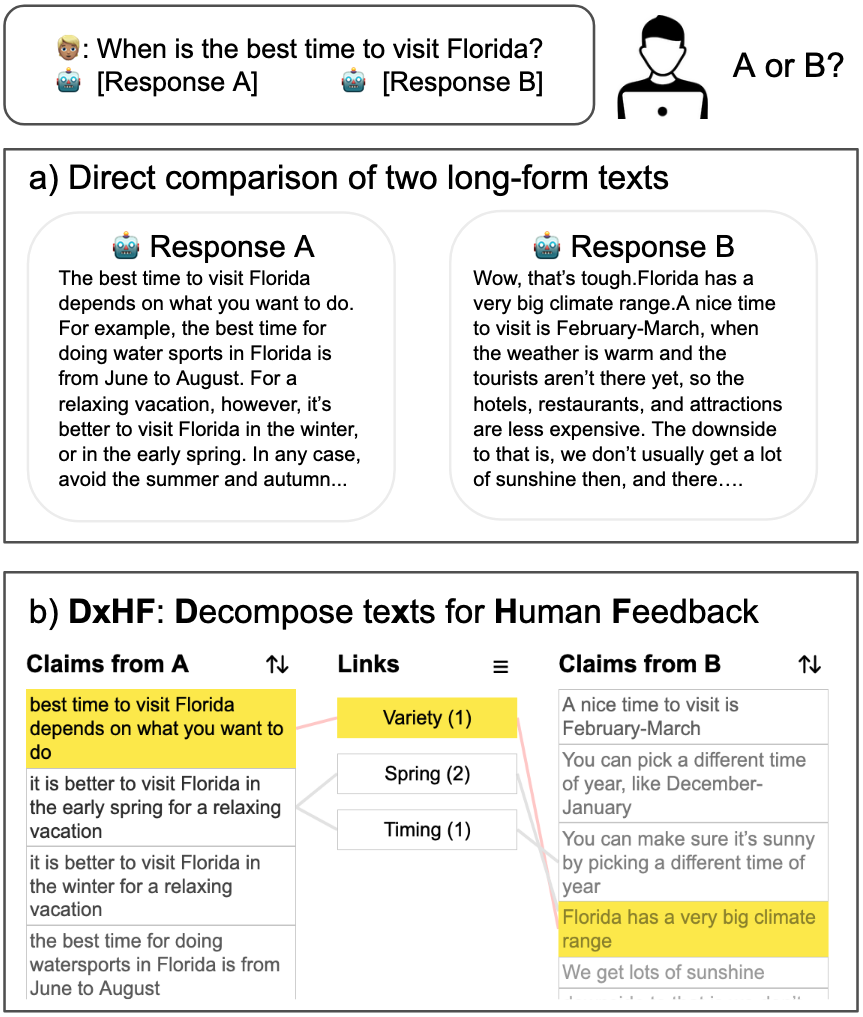}%
\caption{An illustration of the comparison between standard comparative feedback and using \name for human feedback: a) In baseline comparative feedback, annotators must read both long-form texts, memorize mixed information, and mentally identify differences. b) \name breaks down texts into concise, easy-to-read claims. Annotators can pay attention to relevant color-encoded claims and recognize related and connected claims, enabling high-quality feedback. % interactive decomposition
}%
\label{fig:illustration}%
\end{figure}

% For training human-level AI, we need alignment methods that aim to create AI agents that behave in accordance with human intentions~\cite{leike2018scalable}.
% This is increasingly critical for large language models (LLMs) due to their current scale, complexity, and potential impact on society~\cite{ji2023ai}.
Training large language models (LLMs) to align with human values often relies on human feedback, specifically their preferences among different versions of LLM outputs~\cite{leike2018scalable, stiennon2020learning, ouyang2022training}.
One common approach is to present two versions of LLM generations and ask human labelers to select the better one~\cite{bai2022training} (see Figure~\ref{fig:illustration}-a). 
This pairwise human preference data is then used to fine-tune LLMs, such as via reinforcement learning from human feedback (RLHF)~\cite{ouyang2022training, christiano2017deep} or direct preference optimization (DPO)~\cite{rafailov2023direct}.

However, collecting high-quality human feedback on long texts can be challenging.
Compared with short sentences, long texts could exceed the time and attention that readers are willing or able to devote to reading it~\cite{gu2024ai}.
Additionally, long texts tend to have complex text structures and more ambiguous sentences~\cite{zipoli2017unraveling}.
The sentence expression used in the text may also not align well with the reader's familiarity, which can further complicate understanding~\cite{dolch1936basic}.
As a result, providing feedback can be prone to errors, as readers may not carefully read through the text, may find it hard to memorize all the content, and may fail to mentally compare all pieces of information.
\rv{
To the best of our knowledge, there are no existing studies on improving the quality of human feedback for LLM alignment beyond simply comparing two texts~\cite{bai2022training, ouyang2022training}.
}
% To the best of our knowledge, there are no existing studies on user interfaces apart from simply comparing two texts for LLM alignment~\cite{bai2022training, ouyang2022training}.

In this paper, we apply the \textit{decomposition principle} to improve the quality of human feedback for LLM alignment.
The central concept of using the decomposition principle is breaking down the complex task of comparing long-form generated text into simpler comparisons of individual statements. 
% In this paper, we introduce \name ~-- an intelligent user interface designed to support high-quality human feedback for LLM alignment. The central concept is based on the \textit{decomposition principle}~\cite{armstrong1975use}, which breaks down the complex task of comparing long-form generated text into simpler comparisons of individual statements.
This principle allows for more accurate human judgments across a variety of situations~\cite{armstrong1975use}.
% This principle allows for more accurate human judgments across a variety of situations, enabling a more comprehensive consideration of factors compared to making direct comparisons.
Based on this principle, we introduce \name ~-- an intelligent user interface designed to support high-quality human feedback
(see Figure~\ref{fig:illustration}-b), where the long text generation is decomposed into a list of claims. 
% using a fact decomposition technique~\cite{min2023factscore}. 
Each claim contains a single piece of information~\cite{nenkova2004evaluating, min2023factscore}, making it easy to interpret.
To guide the reader's attention to the most relevant parts, 
\name measures the contextual relevance between each claim and the query~\cite{reimers2019sentence} and visually encodes relevance scores using text opacity. 
Less relevant claims are de-emphasized with a lower opacity~\cite{gu2024ai}. 
To facilitate quicker and more effective visual comparisons, 
\name measures the semantic similarity between claims from the response~\cite{devlin2019bert} and links similar claims together with a keyword to summarize the shared meaning in both claims. 
When hovering over each box in the interface, the linked claims and summary keywords are highlighted to help guide attention. 

We conducted three evaluations of \name to understand its effectiveness in improving human feedback quality. 
% First, we performed a technical simulation study using the Boltzmann rational model to evaluate whether text decomposition can enhance accuracy across different levels of human rationality. The result indicates that decomposed claims generally improve the accuracy of feedback, particularly for users with lower rationality.
Firstly, we carried out a technical evaluation to evaluate the decomposition technique, where we simulated synthetic annotators to control how well they gave preference feedback. 
This approach to controlling the level of optimality of a simulated user, called Boltzmann rationality, which is heavily used in human-robot interaction and AI research~\cite{ziebart2010modeling, laidlaw2021boltzmann}, allowed us to identify the potential benefits of the decomposition technique but under conditions we fully control. Simulation results indicate that decomposition improves the accuracy of feedback, specifically in cases where the human annotators are uncertain.
Second, to evaluate how \name facilitates human annotators for comparative feedback, we carried out an online crowdsourcing user study with 160 online participants, utilizing the well-established dataset of human feedback from the most popular votes for machine-generated responses (HH-RLHF~\cite{bai2022training}) as the ground truth. The result indicates that \name improves feedback accuracy by 4.7\% regarding the ground truth ($p<0.05$). Especially, when we consider feedback from people with uncertainty about the answer, the results show a 6.4\% higher accuracy than the baseline ($p<0.01$). Although \name may increase the feedback time, it provides annotators the freedom of control.
Third, we conducted an ablation study with 36 human participants to evaluate the two design features of \name: claim linking and claim ranking. The results indicate that ranking aids in focusing on key information, while linking reduces the effort required for comparison. 
\rv{
While \name is effective for factual or task-oriented comparisons, its design scope is not well-suited for holistic judgments that involve assessing the coherence, tone, or style of text as a whole.
}

In summary, the contribution of this paper is three-fold:
\begin{itemize}[noitemsep,topsep=0pt]
    \item We propose decomposition as a principled approach to presenting text paragraphs for comparison, aiming to improve the quality of human feedback for LLM alignment.
    \item We build a novel user interface \name, designed to decompose and organize text responses for better human feedback. \name is open-sourced at \url{https://sdq.github.io/DxHF}.
    \item We conducted a technical evaluation, a user study, and an ablation study to evaluate \name.
\end{itemize}
\section{Related Work}

This section reviews research studies on current user interfaces for human-AI alignment and relevant work regarding the evaluation of LLMs. Finally, we explore visualization and rendering techniques that inspire our interface design for LLM alignment.

\subsection{User Interfaces for Human-AI Alignment}

\begin{figure}[!t]
\centering

    \begin{subfigure}{\linewidth}
    \centering
    \includegraphics[width=\linewidth]{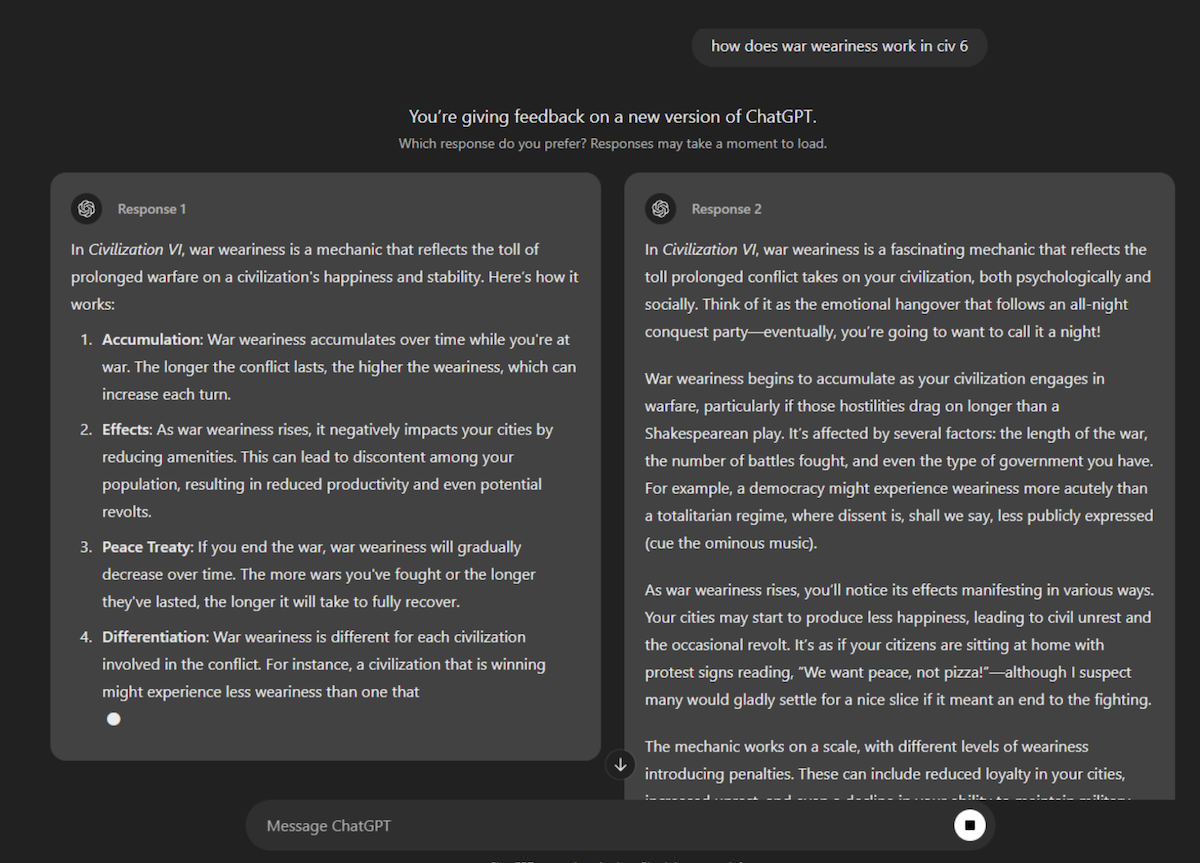}
    \caption{ChatGPT asks casual users to give comparative feedback~\cite{reddit_post}}
    \end{subfigure}
    
    \vspace{1em}
    
    \begin{subfigure}{\linewidth}
    \centering
    \includegraphics[width=\linewidth]{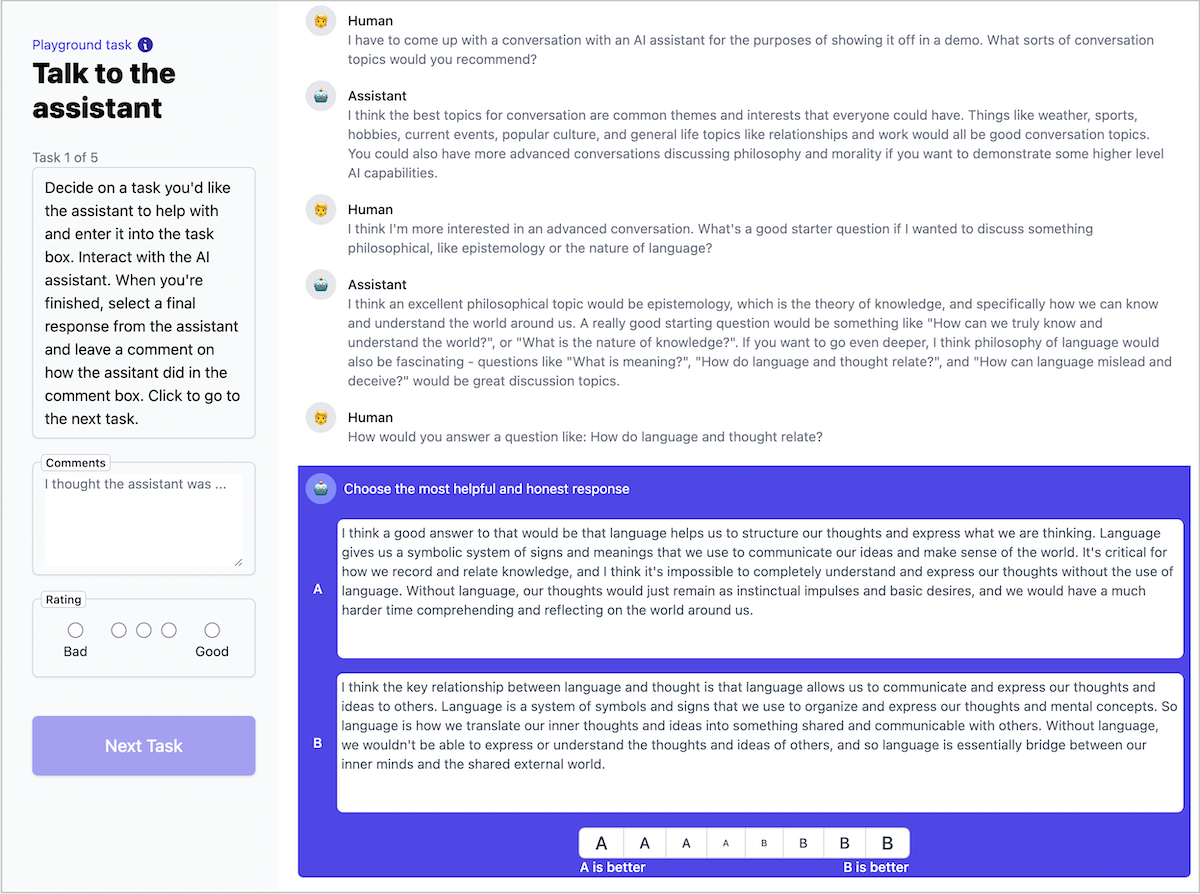}
    \caption{The interface developed by Anthropic for crowdworkers~\cite{bai2022training}}
    \end{subfigure}
  \caption{
  % The screenshots display the standard interface for pairwise comparison used in crowdsourcing human feedback for LLM alignment.
  The screenshots of LLM alignment user interfaces.
  }
  \label{fig:standardUI}
\end{figure}

The classical method for aligning AI with human values typically involves manually designing a well-defined reward function~\cite{mnih2015human}. However, many tasks have goals that are complex and challenging to specify. Learning from human feedback~\cite{christiano2017deep} offers an alternative approach to achieving alignment without the need for manual reward design. This approach begins with the AI agent acting randomly. A human evaluator then continuously observes two behaviors displayed on a user interface and decides which one better aligns with the desired goal. Over time, AI builds a model of the task by modeling the reward that best explains human preferences.
In contrast to the method of pairwise comparisons for gathering preferences, the interactive reward-tuning approach~\cite{shi2024interactive} employs a visualization-based strategy. This allows users to explore the agent's behavior space and discover interesting directions for reward tuning based on the principle of visual information-seeking.
Interactive-RLHF~\cite{jan2025interactiverlhf} improves alignment efficiency by leveraging group-wise comparisons and supports the annotation process with exploratory visualizations.
Additionally, RLHF-Blender~\cite{metz2023rlhf, metz2025reward} provides a user interface that integrates various types of feedback, such as comparisons, demonstrations, and descriptive feedback, to support flexible data collection and improve alignment.
However, most of these approaches are focused on reinforcement learning (RL) agents rather than LLMs.

When aligning LLMs, the behavior of agents is the generated text, which results in user interfaces that allow for comparisons among text. Anthropic built an interface that enables crowdworkers to interact with their models~\cite{bai2022training}, which displays two responses and asks human evaluators to choose which one is more helpful (see Figure~\ref{fig:standardUI}). InstructGPT~\cite{ouyang2022training} and GPT-4~\cite{achiam2023gpt} used a similar approach but introduce two alternative labeling interfaces: (a) a rating page where labelers rate the overall quality on a 1-7 Likert scale and provide various metadata labels, and (b) a ranking page where labelers rank all outputs generated for a specific prompt, thus creating more comparison pairs.
However, these existing user interfaces for LLM alignment highly relys on human reading ability, and face challenges related to user experience, such as inefficient information transfer~\cite{mertsiotaki2025designing}. 
In the AI-assisted writing process, alignment via edits shows~\cite{chakrabarty2024can} exactly where the original text has been modified, allowing for a more detailed comparison, with the edited version often preferred over the original response. However, this approach is limited to text editing when the texts are not significantly different.
Gebreegziabher, Simret Araya, et al.~\cite{gebreegziabher2024supporting} use structural alignment theory to present groups of counterexamples from LLM generation. Based on that, they then introduce a tool, MOCHA~\cite{gebreegziabher2024mocha}, that assists annotators in understanding the alignable differences among data items. However,  MOCHA is designed to annotate data items in batches, not support pairwise comparisons of long-form texts.
Gero, Katy Ilonka, et al.~\cite{gero2024supporting} explored ways to display multiple LLM responses simultaneously and proposed design guidelines to guide future explorations of LLM interfaces. This inspired us to think about how to improve the understanding of LLM responses for high-quality comparative feedback.
\rv{
As opposed to existing approaches, this paper explores a user interface technique that improves the quality of human feedback, specifically focusing on pairwise comparisons for LLM alignment.
}
% This paper introduces a novel user interface that addresses the remaining issues in current user interfaces for LLM alignment. We explicitly focus on obtaining pairwise comparative human feedback for LLM alignment.

\subsection{Interactive Tools for LLM Evaluation}

As LLMs are deployed in more complex and open-ended tasks, evaluation criteria have expanded beyond traditional correctness to encompass more nuanced dimensions of quality and user alignments.  
These evaluations require humans in the loop to provide feedback like perceived helpfulness.
It has created a growing demand for interactive tools that reduce the human workload and speed up the process. 

% Tools like ChainForge~\cite{arawjo2024chainforge} enable users to define criteria by programming or using 

A critical challenge in LLM evaluation is evaluating or judging the LLM output based on human-desired properties that could transform the generated text into structured data, e.g., a true/false value or a score~\cite{shankar2024validates}. 
Arawjo et al. proposed ChainForge~\cite{arawjo2024chainforge}, a visual toolkit that allows users to conduct hypothesis testing on the generated text by inserting programs or defining rules, e.g., whether the outcome contains specific words. 
When evaluating the LLM with high-level criteria, e.g., helpfulness, rule-based evaluators are insufficient. 
A widely adopted alternative is the LLM-as-a-judge approach, where either another LLM or the target model itself is used to conduct the evaluation~\cite{zheng2023judging}.
Kim et al. proposed EvalLLM~\cite{kim2024evallm}, which allows users to specify evaluation criteria in text and apply an LLM to conduct semi-automatic evaluations. 
In LLM Comparator, an interactive visualization tool proposed by Kahng et al.~\cite{kahng2024llm}, the authors use the LLM-as-a-judge approach to support comparisons of the performances of two language models. 

% \begin{itemize}
%     \item Judging llm-as-a-judge with mt-bench and chatbot arena~\cite{zheng2023judging}
%     \item ChainForge: A visual toolkit for prompt engineering and LLM hypothesis testing~\cite{arawjo2024chainforge}
%     \item Who Validates the Validators? Aligning LLM-Assisted Evaluation
% of LLM Outputs with Human Preferences~\cite{shankar2024validates}
%     \item EvalLM: Interactive evaluation of large language model prompts on user-deﬁned criteria~\cite{kim2024evallm}
%     \item Promptfoo: LLM evals \& red teaming
%     \item FactScore: Fine-grained Atomic Evaluation of Factual Precision in Long Form Text Generation~\cite{min2023factscore}
%     \item Loki: An Open-Source Tool for Fact Verification~\cite{li2024loki}
%     \item LLM Comparator: Interactive Analysis of Side-by-Side Evaluation of Large Language Models~\cite{kahng2024llm}
%     \item RELIC: Investigating Large Language Model Responses using Self-Consistency~\cite{cheng2024relic}
% \end{itemize}

Another challenge is deriving and communicating fine-grained insights when evaluating long-form generations, e.g., which part of the text introduces untruthful and unfaithful information. 
Cheng et al. proposed RELIC~\cite{cheng2024relic}, which decomposes the generation into atomic claims~\cite{min2023factscore} and visually annotates the LLM's confidence in these claims in the text. 
The system provides a fine-grained evaluation of the truthfulness and faithfulness of LLM generation and helps users identify untrustworthy content. 
% For example, we know that the generated text contains inaccurate information and 

% The main difference between LLM evaluation approaches and our work is that the LLM evaluation focuses on providing detailed quality analysis of generated results, while 
Compared with existing approaches that conduct holistic evaluation to derive model insights, our work targets a different problem of supporting humans to effectively identify differences and make comparisons between two model generations. 
Our method builds upon and extends atomic claim approaches by decomposing long-form generations into individual atomic claims and visually aligning the relevant ones to facilitate efficient comparison.
\rv{
We believe that other tools can also benefit from the proposed decomposition principle if users require comparative reading. This paper specifically aims to support the LLM alignment task.
}

% the speed of identifying differences and enabling comparisons. 
% Instead of offering extensive explanations, we simplify the information to support quick comparisons for feedback.

\subsection{Visualization Techniques: Text Rendering and Alignment Visualization} % : Text Rendering and Alignment Visualization

Text's visual representation influences audiences' reading efficiency and comprehension correctness~\cite{huth2024eye}. 
Highlighting, e.g., through the background color, bold typeface, or increasing font size, is a common strategy to enhance reading performance. 
Numerous empirical studies have explored what content should be highlighted and how it should be presented to maximize its effectiveness~\cite{joshi2024constrained, strobelt2015guidelines}.
For example, Strobelt et al.'s study shows that font size is the most efficient visual channel for highlighting \cite{strobelt2015guidelines}. 
Joshi and Vogel's study shows that limiting the number of highlighted words enhances reading performance~\cite{joshi2024constrained}. 
Conversely, hiding certain words or letters has also been explored as a technique to support skimming during reading~\cite{gu2024skimmers}.
For example, Gu et al. use font color to encode the saliency of words measured from an LLM-based method: users could skim the light-color words to speed up the reading~\cite{gu2024ai}. 

% Text rendering
% \begin{itemize}
%     \item Constrained Highlighting in a Document Reader can Improve Reading Comprehension~\cite{joshi2024constrained}
%     \item An AI-Resilient Text Rendering Technique for Reading and Skimming Documents~\cite{gu2024ai}
%     \item Why Do Skimmers Perform Better with Grammar-Preserving Text Saliency Modulation (GP-TSM)? Evidence from an Eye Tracking Study~\cite{gu2024skimmers}
%     % \item Eye Tracking on Text Reading with Visual Enhancements~\cite{huth2024eye}
%     \item Visual Foraging of Highlighted Text: An Eye-Tracking Study~\cite{chi2007visual}
% \end{itemize}

Another technique we consider is text alignment visualization, which is used to discover the similarities and differences across two or more text bodies.
Yousef and Janicke categorized seven types of text alignment visualization techniques through a survey~\cite{yousef2020survey}.
Among the seven types of techniques, side-by-side visualization is the commonly used method, which displays the texts next to each other, highlighting aligned units to facilitate direct comparison~\cite{el2017progressive}.
A common limitation of side-by-side visualization is that it requires users to shift their gaze between parallel texts to spot differences~\cite{yousef2020survey}.
So this technique is sometimes combined with aligned barcodes---links that visually associate the aligned text segments~\cite{janicke2017interactive,el2018ltma}.
Remaining techniques, such as text variant graphs that reorganize the text in a diagram~\cite{shankar2024validates}, are either not scalable to long-form text or designed for professional users~\cite{asokarajan2017textile, sevastjanova2023visual}, and thus are out of the scope of this discussion.
% It displays the texts next to each other, highlighting aligned units to facilitate direct comparison.

% Text alignment visualization
% \begin{itemize}
%     \item Interactive Visual Alignment of Medieval Text Versions~\cite{janicke2017interactive}
%     \item TexTile: A Pixel-Based Focus+ Context Tool For Analyzing Variants Across Multiple Text Scales~\cite{asokarajan2017textile}
%     \item Visualizations for text re-use~\cite{janicke2014visualizations}
%     \item A data structure for representing multi-version text online~\cite{shankar2024validates}
% \end{itemize}

% Comparative exploration
% \begin{itemize}
%     \item Progressive Learning of Topic Modeling Parameters~\cite{el2017progressive}
%     \item LTMA: Layered topic matching for the comparative exploration, evaluation, and refinement of topic modeling results~\cite{el2018ltma}
%     \item Visual comparison of text sequences generated by large language models~\cite{sevastjanova2023visual}
%     \item Automatic Histograms: Leveraging Language Models for Text Dataset Exploration~\cite{reif2024automatic}
% \end{itemize}

Our work draws inspiration from the existing visualization techniques. 
Specifically, we combine text rendering to support skimming and a side-by-side alignment visualization to facilitate comparisons among paired claims in two long-form texts.
\rv{Our research focuses on the applications of these techniques in collecting high-quality human feedback and provides studies to understand their effects.
}
% Our research focuses on the applications of these techniques in collecting humans' comparative feedback to LLM generations and provides a quantitative study to understand their effects.

% Text rendering methods can accelerate reading by supporting skimming; Text alignment visualization can help identify similarities and differences, which can facilitate comparisons. 

% Our work draws inspiration from the existing visualization techniques, especially referring to the design spaces of alignment vis and text rendering techniques.

% \begin{itemize}
%     \item Supporting sensemaking of large language model outputs at scale~\cite{gero2024supporting}
%     \item Why Do Skimmers Perform Better with Grammar-Preserving Text Saliency Modulation (GP-TSM)? Evidence from an Eye Tracking Study~\cite{gu2024skimmers}
%     \item Supporting Co-Adaptive Machine Teaching through Human Concept Learning and Cognitive Theories~\cite{gebreegziabher2024supporting}
%     \item Cognition-Inspired Interactive Frameworks for Human-AI Alignment~\cite{gebreegziabher2025cognition}
%     \item "We Need Structured Output": Towards User-centered Constraints on Large Language Model Output~\cite{liu2024we}
%     \item Sensemaking: What is it today?~\cite{russell2024sensemaking}
%     \item Selenite: Scaffolding Online Sensemaking with Comprehensive Overviews Elicited from Large Language Models~\cite{liu2024selenite}
%     \item Tool Support for Knowledge Foraging~\cite{liu2023tool}
% \end{itemize}
\section{Problem Formulation and Objective}

% This section introduces the primary objective of LLM alignment and formulates the problem of comparative feedback for LLM alignment, highlighting the current challenge.

%\subsection{Overall Goal of LLM Alignment}

The overarching goal of LLM alignment is framed in terms of the agent alignment problem: ``\textit{How can we create agents that behave in accordance with the user intentions?}''~\cite{leike2018scalable}.
It can be specifically cast as a sequential decision-making problem where an agent interacts with its environment over time ~\cite{sutton1998reinforcement, hadfield2016cooperative}. At each timestep, the agent takes an action and receives feedback in the form of observations. The objective of alignment is to develop a policy that guides the agent's behavior in line with human intentions, not determined by the environment alone.
Regarding human intentions, researchers have established standards for alignment objectives. The HHH acronym captures key dimensions; an agent should be helpful, honest, and harmless (HHH)~\cite{askell2021general}.
This paper focuses specifically on the aspect of ``helpfulness''.
% as it can be measured in a more objective manner.

\vspace{-0.3em}
\subsection{Problem Formulation}
\label{sec:formulation}

Asking annotators to select from a pair of text generations is the most common method for collecting human feedback for LLM alignment training~\cite{bai2022training, ouyang2022training}.
% Such feedback is easier to collect than asking users to rate each generation individually~\cite{metz2023rlhf}. 
% It involves pairwise comparisons between two options based on human judgments. 
% This approach is widely utilized because it is assumed that it is easier for humans to provide comparative judgments rather than absolute scores~\cite{metz2023rlhf}.
Formally, given a history of human-AI conversational context referred to as a query $Q$, the LLM can generate two response options $A$ and $B$. Each response can consist of long-form text, including a sequence of sentences: $A = \{t^a_1, t^a_2, ..., t^a_n\}$ and $B = \{t^b_1, t^b_2, ..., t^b_m\}$. The user is required to select one of these responses as their preferred option. This comparison result helps tune the model towards better alignment with human values.

Conventionally, human annotators need to read through the two responses $A$ and $B$ using an interface like Figure~\ref{fig:standardUI} for comparison, which involves the following activities:
Feedback is primarily based on the conversation. To give proper feedback, annotators must fully understand the conversational context $Q$.
Annotators are typically asked to read two complete response texts from $t^a_1$ to $t^a_n$ and $t^b_1$ to $t^b_m$.
After reading, people need to identify comparable parts within the texts. For instance, text $t^a_i$ needs to be compared with text $t^b_j$. This may lead them to refer back and forth between the texts.
For each identified difference, people have their own preferences as $\text{p}(t^a_i, t^b_j|Q)$. After comparing multiple differences, people can submit their final preference feedback $\text{Pref}(A,B|Q)$.
% Figure~\ref{fig:standardUI} shows common user interfaces for eliciting comparative feedback with the conversation $Q$ and two generation options $A$ and $B$. 
% The process consists of several steps:
% \begin{itemize}
%     \item[1)] \textit{Understanding the context}. 
%     \item[2)] \textit{Reading sentences in two responses}. 
%     \item[3)] \textit{Identifying differences between texts}. 
%     \item[4)] \textit{Making comparisons}. 
% \end{itemize}

% The preference value is positive if $t^a_i$ is preferred, negative if $t^b_j$ is preferred, or zero if they are considered equal. 
% \begin{equation}
% \text{Pref}(A,B|Q) = 
% \begin{cases}
% A > B, & \text{if } \sum{ \text{p}(t^a_i, t^b_j|Q)} > 0 \\[6pt]
% A < B, & \text{if } \sum{ \text{p}(t^a_i, t^b_j|Q)} < 0
% \end{cases}
% \label{eq:pref}
% \end{equation}
With the collected human feedback data, LLMs can be re-trained or fine-tuned with the methods, such as supervised fine-tuning (SFT)~\cite{taori2023stanford}, reinforcement learning from human feedback (RLHF)~\cite{ouyang2022training, christiano2017deep} or direct preference optimization (DPO)~\cite{rafailov2023direct}

\subsection{\rv{Goal and Design Considerations}} % Objective
% We hypothesize that this process could be facilitated by visualization techniques that help users focus attention on key similarities and differences.
% In contrast, comparing ``walls of text'' is suboptimal for human cognition, because identifying corresponding claims is tedious.
% This can lead to omissions, biases, and generally low-quality feedback (see ~\cite{chen-etal-2024-humans}).
\rv{
The main goal of the paper is to improve the quality of human feedback.
The key challenge in achieving this goal is that is that comparing long text snippets for LLM alignment is cognitively hard, leading to omissions, biases, and generally low-quality feedback (see ~\cite{chen-etal-2024-humans}).
This paper aims to address this challenge by introducing a user interface technique that allows annotators to provide more reliable feedback. To guide and support the design process, we developed a set of considerations that informed our approach.
}

\textit{Reliable feedback across various annotators}. The most central design consideration is to ensure that we collect high-quality human feedback from various human annotators. Unreliable feedback can lead to biased models. Since annotators on the crowdsourcing platform are diverse and contribute varying levels of effort--some providing feedback carefully while others rush through--we aim to design a solution that improves the quality of feedback for both types of human annotators.

\textit{Reasonable feedback time}. The next consideration is to maintain a reasonable time for feedback.  Although there is always a speed-accuracy trade-off~\cite{wickelgren1977speed}, it is important for our method to ensure quality without compromising speed a lot. This will help practitioners collect extensive preference datasets.

\textit{Integration into existing workflow}. The method should be seamlessly integrated into the current feedback workflow. Since human feedback relies on crowdsourcing, it is crucial that the solution is easy to implement on the current web applications and does not require extensive training for annotators to learn how to use it.

%
%Although comparative feedback can be affordable for most people, the human judgements, which LLM alignment highly relies on, could be flawed.
%This is because all the steps outlined in Section~\ref{sec:formulation} are performed mentally by humans.
%This unreliability can be even worse when comparing longer texts, as as they contain more pieces of information that complicate the comparison process.

% ``Humans are unreliable: The human judgements we train against could be flawed: we could miss subtle factual errors, use biased reasoning, or have insufficient context to evaluate the task.''~\cite{chen-etal-2024-humans}

% \begin{itemize}
%     \item Responses generated from human-AI conversations may include several pieces of information.
%     \item Comparing mixed information in two texts is prone to mistakes.
% \end{itemize}
\begin{figure*}
\centering%
\begin{subfigure}{\linewidth}%
\includegraphics[width=\linewidth]{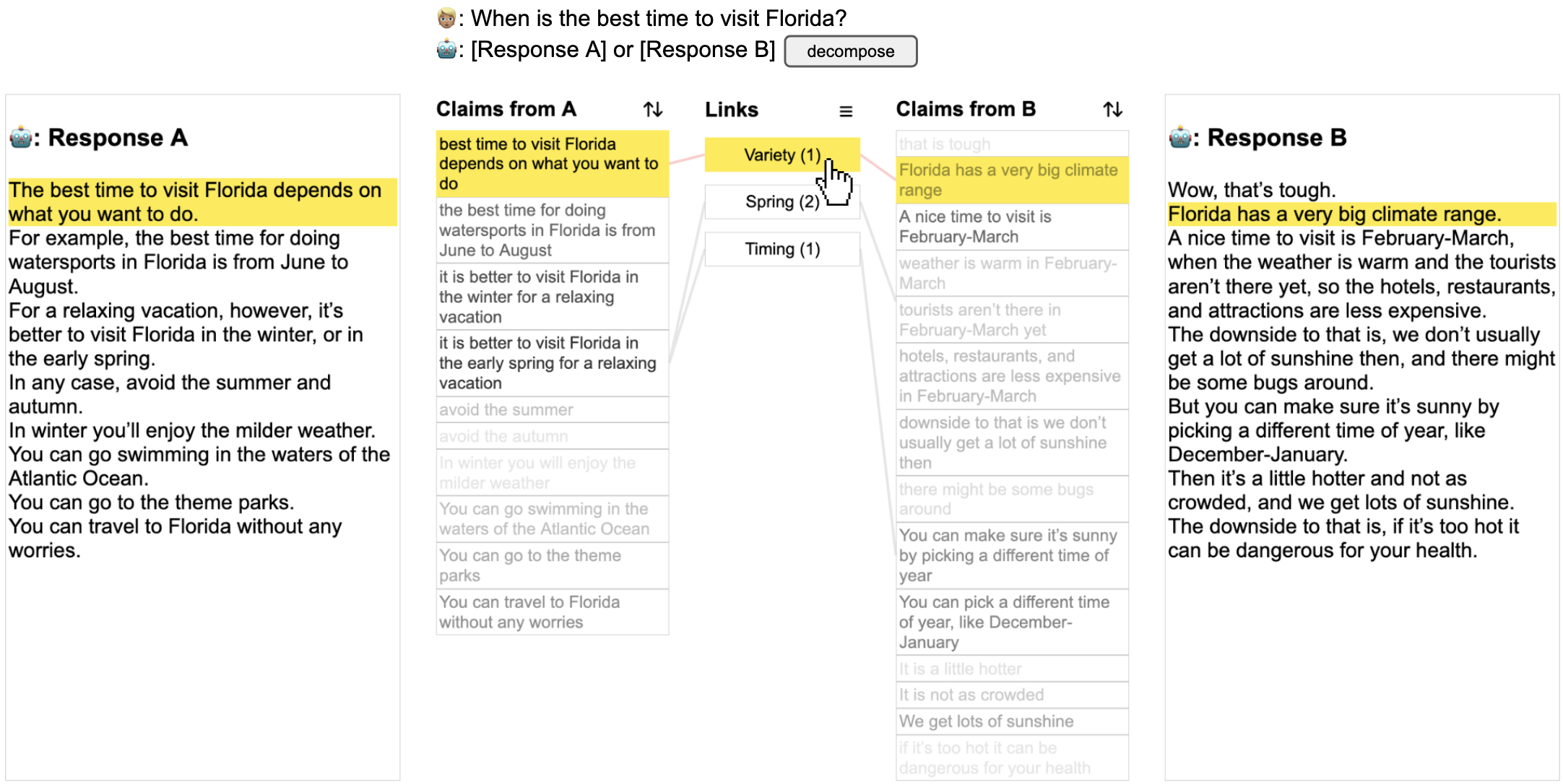}%
\caption{When the user hovers over the keyword of a link, the connected claims are highlighted.}%
\end{subfigure}%
\\%New Line
\begin{subfigure}{\linewidth}%
\includegraphics[width=\linewidth]{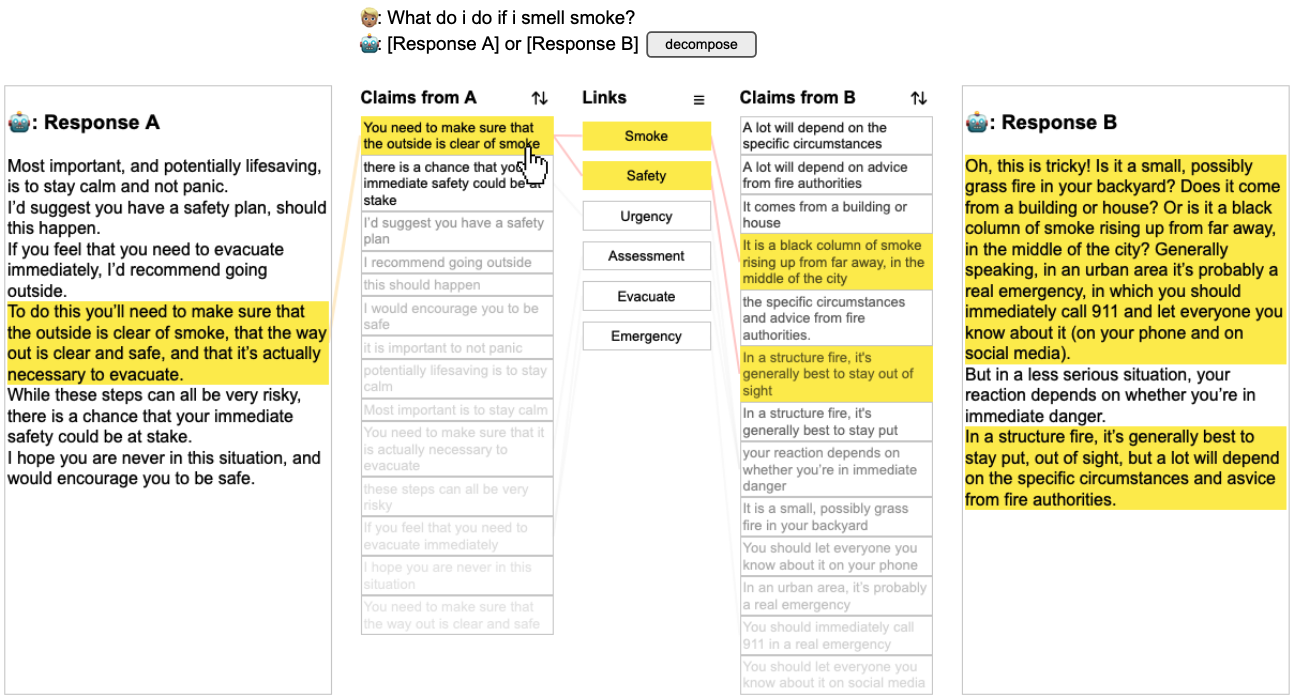}%
\caption{When a user hovers over a claim, all connected links are highlighted along with the corresponding claims on the other side.}%
\end{subfigure}%
\caption{\name decomposes the text and organizes them into individual claims. Similar claims across two responses are connected with a keyword label. By using hover highlights, human annotators can more easily identify differences and compare the claims. We provide two examples: (a) grouping the links ($\equiv$) and hovering over a keyword, and (b) sorting the claims ($\uparrow\downarrow$) and hovering over a specific claim.}%
\label{fig:ui}%
\end{figure*}

\section{\name: Interactive Decomposition}

% To address the challenge of collecting reliable human feedback for LLM alignment, this section presents the design of our proposed user interface.
% We present our motivating design considerations, followed by a design overview, and a detailed explanation of the computational workflow and user interface design.

We design \name to tackle the challenge of collecting high-quality human feedback for LLM alignment. 
This section presents the decomposition principle used by \name, followed by a description of the computational workflow and user interface design.

\subsection{Decomposition Principle}

% The primary goal is to improve the quality of human feedback.
% The main design principle we used to achieve the goal is the \textit{decomposition principle}, as established by J. Scott Armstrong in 1975~\cite{armstrong1975use}. 
\rv{
The main design principle we used to improve the quality of human feedback is decomposition.
Why focus on decomposition? The \textit{decomposition principle}, as established by J. Scott Armstrong~\cite{armstrong1975use}, has demonstrated that it allows for a more thorough consideration of factors compared to making direct judgments.
}
% This principle has been shown to allow for a more comprehensive consideration of factors compared to making direct judgments.
Specifically, utilizing the decomposition principle often results in more accurate human judgments in a variety of situations. Its value is particularly evident in supporting robust decision-making when uncertainty is high, which frequently occurs when collecting feedback on crowdsourcing platforms due to human annotators having limited attention spans.
Furthermore, analyzing information in decomposed components through computational methods can be more effective than relying solely on mental processing.

In our case of comparative feedback for LLM alignment, we decompose long-form generated texts into a series of straightforward statements, which we refer to as atomic claims. Each claim conveys a single piece of information in a simple format.
\rv{Research has demonstrated that long-form texts can be verified through decomposition into atomic claims~\cite{fan-etal-2020-generating, wright-etal-2022-generating}.}
When we ask annotators to read these claims instead of the original text, we also take into account the \textit{integrity principle} during the decomposition. This principle emphasizes the importance of preserving the original information as much as possible while reorganizing it to facilitate comparison. This method is commonly used in text rendering~\cite{gu2024ai} and visualization~\cite{sultanum2018doccurate} to ensure fidelity to the original text.

To effectively organize and present the decomposed claims for comparative feedback, 
\rv{we guide annotators' attention by applying the visualization principle, known as \textit{Focus+Context}~\cite{card1999readings}. This approach allows annotators to concentrate on the most relevant parts of the information while simultaneously presenting the surrounding context.}
% we base our user interface design on the principle of information visualization: \textit{Focus+Context}~\cite{card1999readings}.
The main concept behind Focus+Context is to allow annotators to concentrate on the most relevant parts of the information while simultaneously displaying surrounding context. In our application, this means that annotators can view the key claims without losing sight of the original text and the relevant information needed for comparison.

Based on these principles, our design primarily modifies the three main steps (2-4) in the comparison process outlined in Section~\ref{sec:formulation}, ensuring that it can be easily integrated into existing applications without significantly altering the workflow:
Instead of reading two texts $A$ and $B$, people are asked to read through two lists of claims from $c^a_1$ to $c^a_N$ and $c^b_1$ to $c^b_M$. They can choose to allocate their effort selectively, reading some claims carefully while skimming others;
During reading, people are provided with corresponding information on both sides, indicating whether there are corresponding claims in the other list. The identification of differences can be done during the reading process;
Rather than comparing mixed information within sentences, they can now directly compare two claims, $\text{p}(c^a_i, c^b_j|Q)$, when there is a comparable pair, or $\text{p}(c^a_i, *|Q)$ if there is no comparable information available on the other side. The comparisons can be much simpler as there is only one piece of information in each claim.

\subsection{Computational Workflow}
\label{sec:decomposition}

We introduce the computational workflow for interactive decomposition. To support interactive comparative analysis, we present ranking and linking techniques after decomposition.

\paragraph{Decomposition}

The response texts consists of a set of sentences ($A = \{t^a_1, t^a_2, ..., t^a_n\}$ and $B = \{t^b_1, t^b_2, ..., t^b_m\}$). These sentences can be segmented from the text using the NLTK sentence tokenizing feature~\cite{bird2006nltk}. However, each sentence might still have a complex structure and contain mixed pieces of information, burdening the people's understanding.
To address this, our approach further breaks down each sentence $t_i$ into individual claims $\{c_1, ..., c_j\}$, where each claim $c_i$ is a concise sentence that conveys only one piece of information. 
For instance, the sentence ``\textit{In addition to his acting roles, he has written and directed two short films and is currently in development on his feature debut.}'' can be decomposed into 1) ``\textit{He has acting roles}''; 2) ``\textit{He has written two short films}''; 3) ``\textit{He has directed two short films}''; 4) ``\textit{He is currently in development on his feature debut}''.
Therefore,  this decomposition creates a claim list for the responses: $A = \{c^a_1, c^a_2, ..., c^a_N\}$ and $B = \{c^b_1, c^b_2, ..., c^b_M\}$.
The decomposition approach is powered by LLMs. We feed each sentence to GPT4 with a series of instructions to break it down into a collection of atomic claims~\cite{min2023factscore}. At the same time, we aim to use only the words found in the original text to ensure fidelity to the source material. 
We follow a similar strategy in reading~\cite{gu2024ai} by modifying the prompt to prevent the addition of new words. While the model may occasionally add or modify some words, this does not significantly affect the main comparison. The full prompt fed to the model is attached in Appendix~\ref{sec:decompose_prompt}.

\paragraph{Ranking}
Each claim holds a different level of importance within the overall text, with some being more significant than others. To address this, we can rank all claims based on their relevance to help annotators focus their attention on the more important points.
In this context, importance refers to a claim's contribution to answering the questions posed during the conversation. We measure this relevance as a way to rank the claims. Given the conversation, which we refer to as a query $Q$, we utilize a Cross-Encoder architecture~\cite{reimers2019sentence} to measure the contextual relevance of each claim: $r_i = \text{CrossEncoder}(Q, c_i)$. The value $r_i$ ranges from 0 to 1, where 0 indicates irrelevance and 1 is full relevance.

\paragraph{Linking}
Given two lists of claims, $A = \{c^a_1, c^a_2, ..., c^a_N\}$ and $B = \{c^b_1, c^b_2, ..., c^b_M\}$, we can analyze the similarity between these lists to assist in comparing two texts. We connect similar claims from both lists (for example, $c^a_i \leftrightarrow c^b_j$) to enable annotators to quickly reference specific parts and identify differences. 
Specifically, we link pairs of claims based on their semantic textual similarity, which we determine using the cosine similarity of their textual embeddings~\cite{devlin2019bert}. 
To improve the speed and effectiveness of comparisons, we provide a keyword that summarizes the meaning of the linked claims. This is powered by prompting GPT4 (see prompt in Appendix~\ref{sec:keyword_prompt}). Each keyword is used as a label to represent the pair of two linked claims.

\subsection{User Interface Design}
\label{sec:interface}

The user interface of \name consists of three main areas (see Figure~\ref{fig:ui}): two vertical lists of claims derived from different responses, along with a connecting links region in the middle.
The claim lists are positioned near the original text, with lines and keyword labels linking related claims across the two lists. The processed data from Section~\ref{sec:decomposition} serves as the basis for the \name user interface. We avoided using complex visualization techniques to minimize the learning curve, as the tool is designed for crowdsourcing use. 

To reduce visual clutter, links with the same keywords can be grouped and bundled together by clicking the $\equiv$ button at the top of the link keywords. As shown in Figure~\ref{fig:ui}-a, the middle keyword list displays the number of links related to specific keywords. For instance, in this example, there are two links connected to the topic ``Spring''.
In each list of claims, the opacity of the text indicates their relevance. We chose opacity over alternatives like background color or stylistic indicators to minimize visual distraction, similar to prior studies~\cite{10.1145/1753846.1754093, gu2024ai}. Less relevant claims are de-emphasized by using a lighter shade of gray text that maintains legibility~\cite{gu2024ai}. The claims are organized in the same narrative order as the original text but can also be ranked by their relevance to the conversation by clicking the $\uparrow\downarrow$ button at the top of the claim list. As shown in Figure~\ref{fig:ui}-b, the claims from both sides are arranged according to their relevance, with font colors ranging from dark to light.
We discuss the alternative designs in Appendix~\ref{sec:alternative_designs}.

\name uses hovering interaction to support Focus+Context principle.
When hovering over each item, the corresponding parts will also be highlighted for reference. For example, when you hover over the keyword of a link (see Figure~\ref{fig:ui}-a), the two claims on either side will be highlighted, along with their original parts in the text responses. Additionally, when you hover over a claim (see Figure~\ref{fig:ui}-b), all the links associated with that claim will be highlighted, as well as the corresponding claims on the opposite side. 

%Tino: Always keep figures floating, do not use [h].
\begin{figure}%
\centering%
\includegraphics[width=\linewidth]{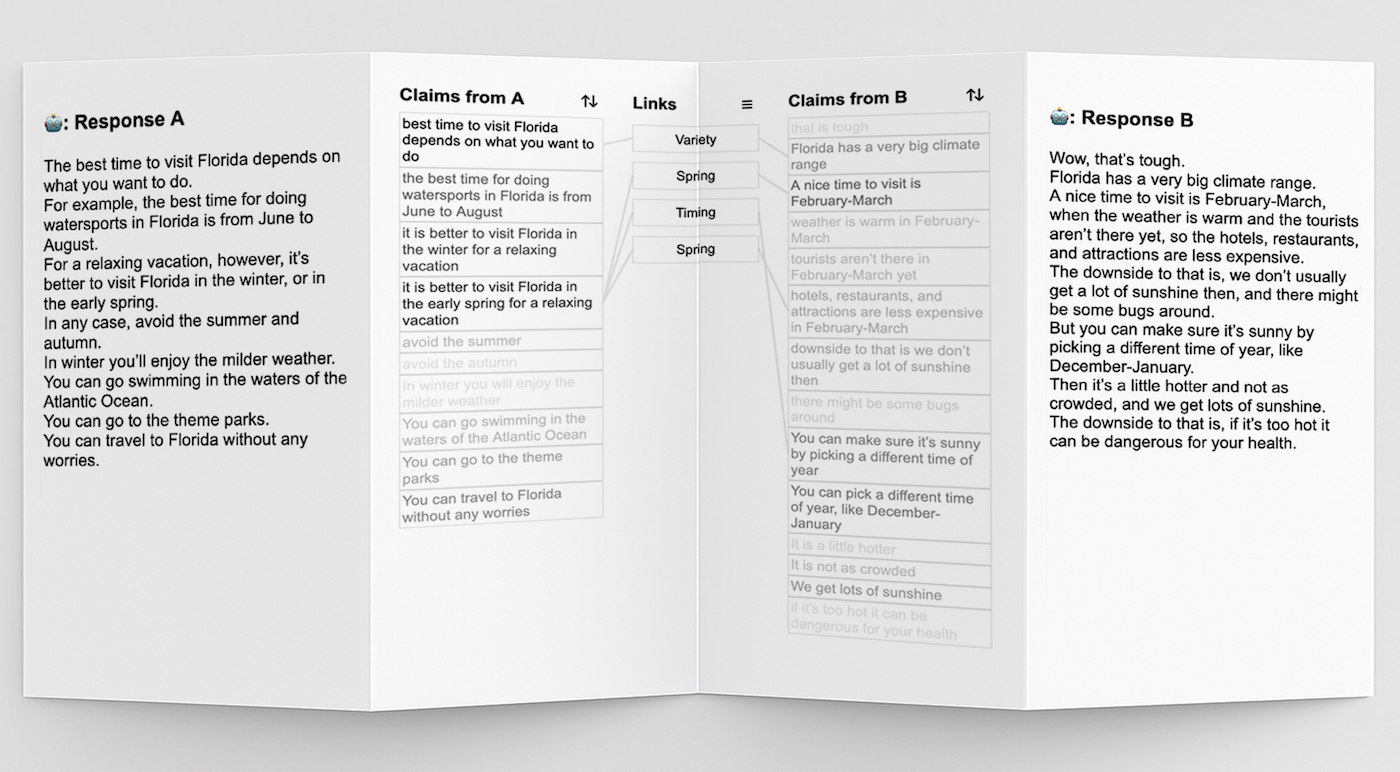}%
\caption{The design is based on a visual metaphor of an accordion fold, allowing human annotators to flexibly fold for reading text or unfold to analyze decompositions.
}%
\label{fig:metaphor}%
\end{figure}

To facilitate seamless integration with existing comparison methods, the \name interface extends the pairwise comparison feature using the metaphor of an accordion fold (see Figure~\ref{fig:metaphor}). Annotators have the freedom to decide to fold to only read full text or unfold to see detail decompositions. 
In our implementation, we include a ``decompose'' button located next to the preference query, allowing easy access to the \name interface. When a user clicks the ``decompose''  button, the two text responses will animate and shift to opposite sides of the screen. In the center, the decomposed claims, along with additional supporting information, will be displayed. This feature enables annotators to select \name based on the complexity of their comparison tasks. The \name interface is developed using D3.js~\cite{bostock2011d3}, which ensures straightforward integration into any web-based application. 
The demonstration of the interaction can be found in the Supplemental Video.
\section{Technical Evaluation}
\label{sec:simulation}

We conduct a simulation study to evaluate the effectiveness of the proposed decomposition methods in supporting text comparison. 
Specifically, we use LLM-based synthetic participants~\cite{argyle2023out} to simulate human feedback.
There are two main advantages for using simulations as a method of technical evaluation. 
First, user simulation allows us to study the conditions in which \name can be beneficial to users, providing a broad foundation before moving on to an empirical study. 
This approach is gaining popularity in research~\cite{syntheticusers}. While we do not assume that an LLM can replicate the nuance of human behavior, they are trained on a vast corpus of human language, which can help us gain insights into tendencies in human judgments~\cite{dillion2023can}.
Second, user simulation enables us to test our hypotheses with different levels of user capability, i.e., rationality, which is challenging to control in traditional empirical studies. It allows us to observe how users with varying levels of ability interact with the techniques. The level of ability here basically means the human's ability to respond to the task according to their true preference.
This type of user simulation is commonly employed in human-robot interaction and AI research~\cite{ziebart2010modeling, laidlaw2021boltzmann}.

\subsection{Simulated Annotators}

We built simulated annotators by modeling their strategies in comparative feedback and their preferences.
The simulated annotator is presented with the task of selecting which of two generated texts is more helpful, given the conversation as the context.
We use the LLM-as-a-judge method~\cite{gu2024survey} to simulate how people judge a text, as the LLM-as-a-judge approach has already demonstrated a strong similarity with humans~\cite{liu2023g}. 
In practice, we 1) use LLMs to score both texts for simulating human judgement of the texts, and then 2) use these scores to simulate human comparative feedback. Below, we introduce the technique details about these two steps.

\paragraph{Simulated Human Judgement}
Practically, we prompt an LLM to measure the helpfulness of a given text on a scale from 0 to 1 (see Appendix~\ref{sec:synthetic_prompt} for the full prompt).
We build four simulated annotators with different strategies:
\begin{itemize}
    \item \textit{Baseline}: This strategy is to directly compare two text responses, where we can directly get $s(A)$ and $s(B)$ by feeding the full text to the LLM.
    \item \textit{Decomposition}: This strategy focuses on comparing decomposed claims rather than long-form text. For the strategies that involve decomposition, we compute the average preference score for each response: \( s(A) = \sum_{j=1}^{N} \text{s}(c^a_i) \) and \( s(B) = \sum_{j=1}^{M} \text{s}(c^b_j) \).
    \item \textit{Decomposition + Ranking}: This strategy is similar with Decomposition. The difference is that it takes into account the relevance ranking of the decomposed claims. Only comparisons are made for claims that exceed a certain relevance threshold $th$: \( s(A) = \sum_{r_i\ge th} \text{s}(c^a_i) \) and \( s(B) = \sum_{r_j\ge th} \text{s}(c^b_j) \). We set the threshold $th$ of relevance at 0.3 to filter out less relevant claims.
    \item \textit{Decomposition + Ranking + Linking}: This strategy considers both the relevance ranking and similarity linking of the decomposed claims. It compares claims that surpass a specified relevance threshold as well as their linked claims: \( s(A) = \sum_{r_i\ge th} \text{s}(a_i) + \sum_{c \in L} \text{s}(c^a_i) \) and \( s(B) = \sum_{r_j\ge th} \text{s}(c^b_j) + \sum_{c \in L} \text{s}(c^b_i) \). We include links with similarity at 0.7 in $L$ to ensure linking high similarity of linked claims.
\end{itemize}

\paragraph{Simulated Human Feedback}
The final preference is modeled using the Boltzmann rational distribution~\cite{ziebart2010modeling, laidlaw2021boltzmann}, where rationality indicates the extent to which feedback aligns with true preferences. It follows the equation below:
\begin{equation}
P(A \text{ is preferred over } B) = \frac{\exp(\beta \, s(A))}{\exp(\beta \, s(A)) + \exp(\beta \, s(B))}
\end{equation}
% where $\beta\in [0, \infty)$ defines the level of rationality. 
% (0 means fully stochastic and $\infty$ means fully rational).
\rv{where $\beta \in [0, \infty)$ defines the level of rationality, representing how consistently a simulated annotator selects the better response. A lower $\beta$ indicates more noisy behavior and a higher one  indicates more rational behavior.  We adjust $\beta$ to simulate various annotators for modeling their decision-making processes.}

\subsection{Ground-truth Data}

% We use the human feedback dataset HH-RLHF~\cite{bai2022training} as our ground truth, which is based on the most popular votes. 
\rv{HH-RLHF~\cite{bai2022training} is a frequently used dataset for evaluating human preferences regarding the helpfulness of human-AI dialogues. It contains human preference labels for over 16,000 comparison pairs of user-model conversations collected by Anthropic~\cite{hh-rlhf}.}
% We use the HH-RLHF dataset~\cite{bai2022training} created by Anthropic, which contains human preference labels for over 16,000 comparison pairs of user-model conversations. 
From this dataset, we consider one- and two-round conversations and select a subset of the ``non-trivial'' comparison instances based on the response length. 
Specially, we exclude the short responses (fewer than five sentences) 
as well as pairs with large length discrepancies (differences greater than 30 words) to mitigate potential human biases favoring more verbose answers~\cite{park2024disentangling}.
Additionally, we remove instances that require domain-specific knowledge to evaluate, such as those involve cooking recipes, identified using keywords. 
Finally, we sample 50 instances from the remaining dataset.  
% Besides, we also exclude 
% From this dataset, we filter a subset for our study. The filtering criteria include:
% 1) \textit{Short conversation history}. We limit the conversation to a maximum of 2 rounds, as we want to avoid lengthy interactions.
% 2) \textit{Similar length of the responses}. The difference in length between the two responses must not exceed 30 words. This ensures that significant disparities in response length do not influence human judgment~\cite{park2024disentangling}.
% 3) \textit{Informative responses}. Each response should contain more than 5 sentences. Our focus is on the approach's performance with longer responses.
% 4) \textit{Removal of cooking-related conversations}. Besides, we also removed all cooking-related conversations, which might be too step-by-step routine, by filtering out keywords like ``cook'', ``food'', ``recipes'', etc.
% % To ensure that these preferences from the HH-RLHF dataset can be used as ground truth, the authors double-check the preferences manually to confirm that the preferred responses are actually helpful. 
% It leads to 50 instances in total at the end.

\subsection{Results}

\begin{figure}
\centering%
\includegraphics[width=\linewidth]{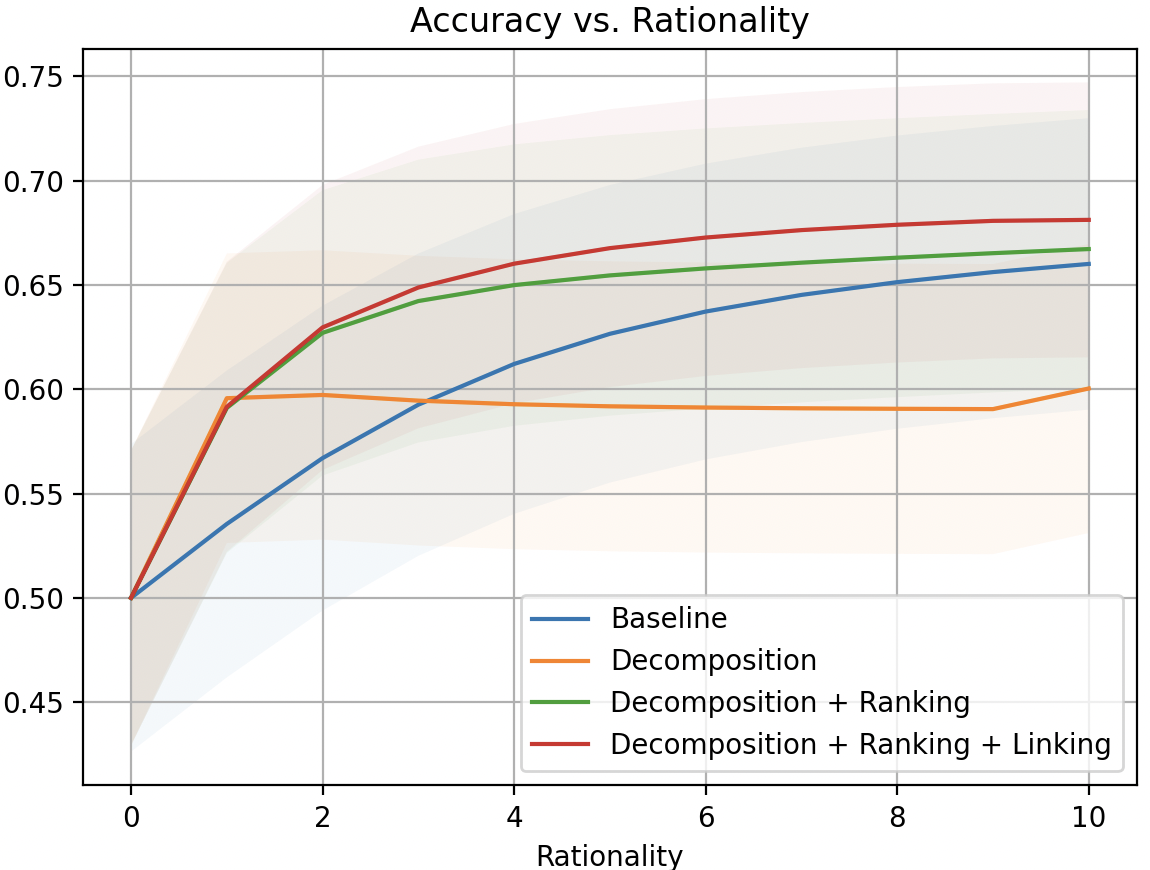}%
\caption{Accuracy vs. rationality of simulated annotators, where rationality means how strongly an annotator's feedback is guided by the true preference, rationality=0 is stochastic and rationality=10 is close to fully rational. Decomposition with ranking and linking is the most accurate strategy across all rationalities. 
% When the level of rationality is below 3, all decomposition methods outperform the baseline.
}%
\label{fig:Simulation}%
\end{figure}

% The results indicate that decomposition can make high-quality human feedback. 
\rv{
This simulation study theoretically confirms the advantages of the decomposition techniques employed by \name.
As shown in Figure~\ref{fig:Simulation}, the performance of the strategies improves as rationality increases, ranging from 0 (completely random) to 10 (nearly fully rational).
The findings from this study demonstrate that implementing decomposition with appropriate ranking and linking techniques generally improves the accuracy of feedback across all levels of rationality.
The combination of decomposition with ranking and linking is the most effective approach, while decomposition with ranking alone ranks second. 
When the level of rationality is below 2, all decomposition methods outperform the baseline approach. This supports our hypothesis based on the decomposition principle~\cite{armstrong1975use}, which suggests that decomposition enhances robust decision-making in low-rationality conditions.
However, there is a plateauing effect of the decomposition method observed when $\beta$ above 3. This occurs because the simulators treat all comparisons of claims equally, which increases the overall noise in comparisons. Although simulations can benefit from controlling the rationality level, they are limited to modeling how users actually read, as users tend to selectively focus on certain comparisons. This led us to conduct the crowd-sourcing user study.
}
% Implementing decomposition with appropriate ranking and linking techniques generally improves the accuracy of feedback across all levels of rationality. The findings from this study demonstrate that the combination of decomposition with ranking and linking is the most effective approach, while decomposition with ranking alone ranks second. 
% This simulation study theoretically confirms the advantages of the decomposition techniques employed by \name.

% \section{Empirical User Studies}

% Following the technical evaluation, we conducted two empirical user studies using the actual \name user interface to evaluate how it facilitates human annotators for comparative feedback. 
% The first crowdsourcing user study focused on comparing \name with the baseline user interface. The second ablation study focused on evaluating the feature designs of \name.

\section{User Study}
\label{sec:userstudy}

In our technical evaluation, we focused on the decomposition techniques. Following this, we conducted an empirical user study using the actual \name user interface to assess how effectively it supports human annotators in providing comparative feedback. 
Specifically, we carried out a crowdsourcing user study to determine whether \name improves the quality of human feedback compared to a baseline tool.
There are two main reasons to involve a crowdsourcing user study to evaluate \name. First, crowdsourcing is the most common method for practitioners to collect human feedback for LLM alignment. We can test the behavior of crowdworkers in an ecologically valid condition that closely resembles the true context in which RLHF is done~\cite{ouyang2022training, bai2022training}. Second, a crowdsourcing user study can include a diverse range of participants in a short term~\cite{kittur2008crowdsourcing}, allowing for the testing \name with different styles of annotators.

\subsection{Experimental Design}

Our evaluation followed a within-subjects design, where each participant used both the baseline and \name. 

\paragraph{Participants}
We recruited 160 general participants from Prolific (80 females, average age 34.1, $SD=8.6$), all of whom are native English speakers. Out of these participants, 155 had prior experience using ChatGPT. They reported an average of 1.8 years of experience with ChatGPT-like tools (\(SD = 1.8\)). Participants were required to complete the task on a laptop or desktop computer. Those who provided feedback too rapidly (less than 10 seconds per response) were excluded from the study. Each study session lasted approximately 30 minutes, and participants were compensated £6 upon completion.

\paragraph{User Interfaces}
Participants interacted with two user interfaces: \name and \textit{baseline}. In the \textit{baseline} interface, participants manually compared text responses as in the default setting of ChatGPT. We designed the \textit{baseline} interface to match the visual style of \name. 
% With \name, participants accessed the full functionality of that user interface for comparison. 
To minimize any potential ordering effects, we counterbalanced the presentation order of the two user interfaces.

\paragraph{Data}
We reused and filtered the ground-truth data from our simulation study (Section \ref{sec:simulation}), removing conversations requiring professional knowledge (e.g., ``\textit{How does a rock transform into a crystal?}'' or ``\textit{What is a logical fallacy?}''). This resulted in 10 tasks, which will be divided into 5 tasks per tool for each study session.

\paragraph{Procedure}
Each participant completed two sessions, each with a different user interface and five comparison tasks.
Each session began with a tutorial and a demonstration using an example that was not used for the actual comparisons. 
For each comparison, we asked the participants which response they found more helpful in the context of the conversation. They were also asked to rate their confidence in their responses on a Likert scale from 1 to 5. 
The study interface can be found in Appendix~\ref{sec:study_interface}.
During the session, we recorded the completion time for each task.
All sessions were conducted on the participants' own desktops, ensuring a similar setting of human feedback for LLM alignment.
After finishing both sessions, participants were asked to leave comments on both interfaces. Finally, they completed an exit questionnaire, which collected general information (such as age and gender) and their experiences with AI tools like ChatGPT.

\subsection{Results}

We analyze the study results based on feedback accuracy, completion time logs, and comments.

\begin{figure}[!t]
\centering
    \begin{subfigure}{\linewidth}
    \centering
    \includegraphics[width=\linewidth]{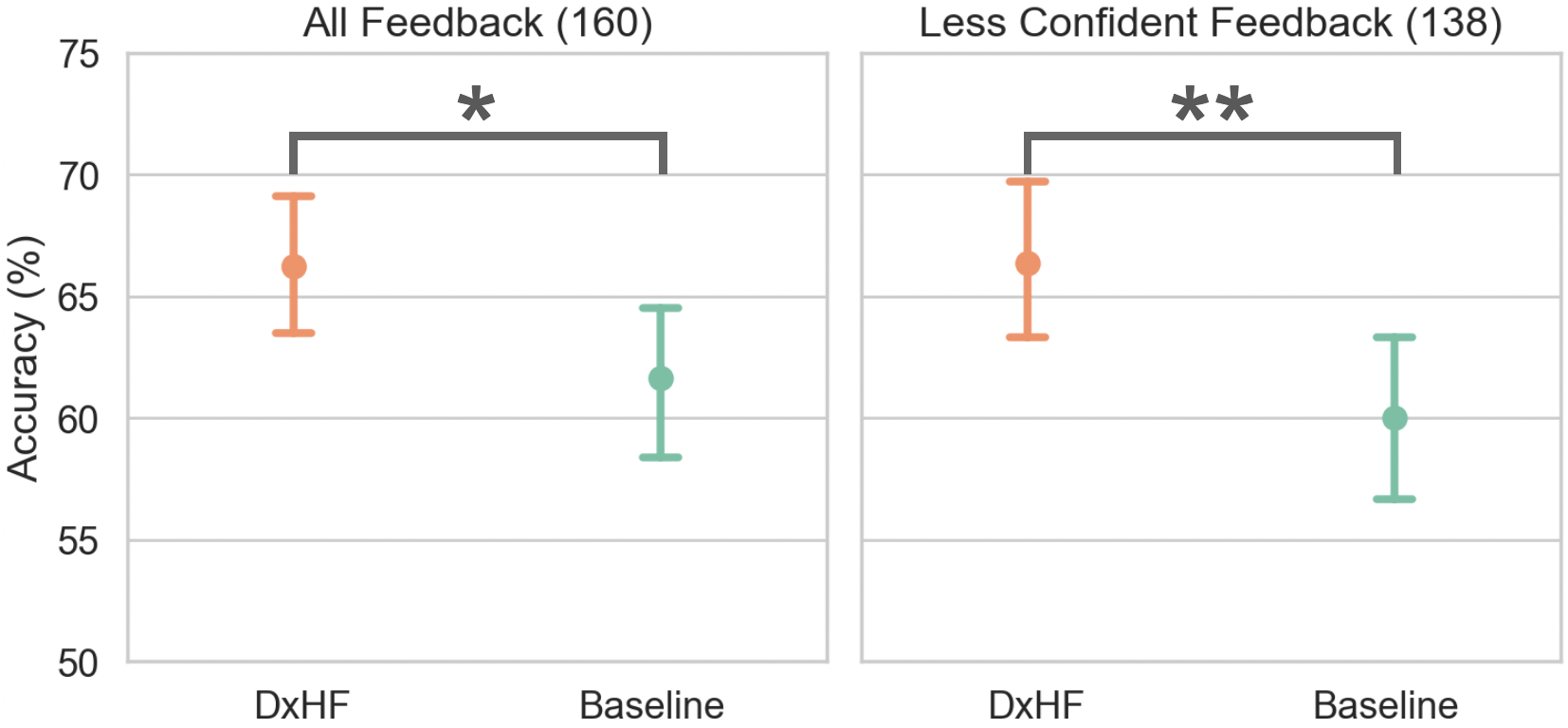}
    \caption{\name is 4.7\% more accurate than the Baseline, and perform significantly better (6.4\%) in cases of less certainty. The error bars represent 95\% confidence interval. (*: $p<0.05$, **: $p<0.01$)}
    \end{subfigure}
    
    \vspace{1em}
    
    \begin{subfigure}{\linewidth}
    \centering
    \includegraphics[width=\linewidth]{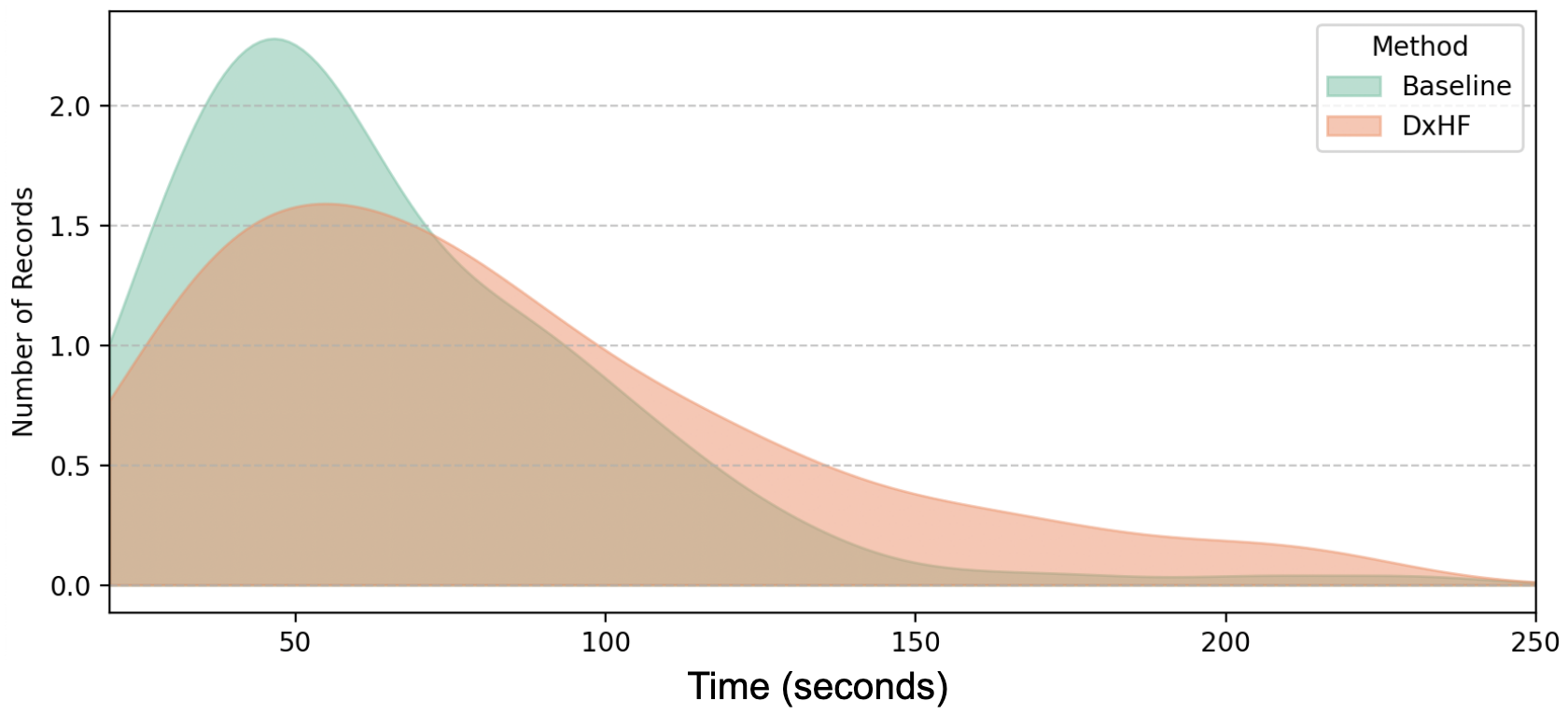}
    \caption{\name spends more time than the baseline, but most of the cases (95\%) are less than 200 seconds.}
    \end{subfigure}
  \caption{Accuracy and speed analysis of the crowdsourcing user study results}
  \label{fig:study}
\end{figure}

\paragraph{Accuracy and speed}
The comparison of feedback accuracy and speed is presented in Figure~\ref{fig:study}. 
Regarding accuracy (see Figure~\ref{fig:study}-a), the use of \name (66.3\%) demonstrates greater accuracy than the baseline approach (61.6\%), as indicated by the statistically significant results from the non-parametric Wilcoxon signed-rank test~\cite{woolson2005wilcoxon} ($Z = 2827.5, p = 0.0293$). When we focus solely on participants who expressed less certainty—excluding those who rated their certainty as 5—we are left with 138 participants. In this subset, the feedback accuracy of \name is highly significantly greater than that of the baseline approach ($Z = 1910, p = 0.0059$). This finding aligns with the simulation study in Section~\ref{sec:simulation}, supporting the point that \name assists individuals who are uncertain in their responses, allowing them to express their preferences more effectively.
In terms of speed, we logged the average feedback time. As shown in Figure~\ref{fig:study}-b, using \name takes more time compared to the baseline, which is expected since participants may incur some learning costs while using the tool and spend additional time analyzing its features. Specifically, \name results in an average increase of 18 seconds in feedback time. 95\% of feedback sessions are completed in under 200 seconds. In practice, practitioners can decide on the trade-off between accuracy and speed based on their specific needs.

\paragraph{Participant Comments}
Participant comments were broadly positive toward \name, highlighting ease of identifying key information (e.g., ``I found [\name] very helpful! The supporting claims made it easier to break down and compare the logic and relevance of each response.'', ``[\name] makes the comparison easier and faster, as the main points are well mapped to access information.'' and ``[baseline] was all my own personal interpretation. [\name] was with highlights of key potential information to take into account.'', \name helped highlight important details and made making judgments easier and quicker.), improved confidence in decisions (``[\name] was very helpful, I thought it helped me make better informed decisions.'' and ``[\name] helped me spot subtle differences more clearly and feel more confident in my choice.''), and reduced cognitive load during complex comparisons (``[\name] had a better tool of highlighting things, while in [baseline] I just did that manually in my head.''). 
Participants who preferred the baseline liked the straightforward way for comparison (e.g., ``[Baseline] was straightforward and allowed intuitive comparison based on personal judgment.'' and ``[Baseline] was straight to the point''). Some mention that [\name] may introduce unnecessary complexity for simple comparisons, such as ``I found [baseline] much easier to compare because I wasn't so distracted by the extra information'' and ``[\name] provided more details but seemed a bit more confusing given all the information it provided.''.

\paragraph{Balancing Complexity and Depth}
Participants noted distinct preferences for using \name and the baseline tool depending on task complexity (e.g., ``[baseline] was much easier, simpler and direct to use. [\name] went further in the reasoning and explanations. I think I would use [\name] for complex questions and [baseline] for more simple general questions.'' and ``Both tools were useful in their own way. [baseline] was straightforward and allowed for intuitive comparison based on my personal judgment. [\name] added a helpful layer of supporting information, especially the hover-to-highlight feature, which made it easier to understand the relevance of specific claims in each response. However, it also made the task slightly more complex.''). Many participants appreciated \name for complex questions that required in-depth analytical thinking, emphasizing its ability to clearly highlight relevant information and relationships between responses. One participant remarked, ``[\name] really helped with more complicated queries; it broke down responses clearly, allowing me to compare subtleties.'' 
Conversely, for simpler, general questions, participants favored the directness of the baseline tool, stating that it provided straightforward insights without unnecessary detail: ``For simpler tasks, the baseline was faster and easier—I didn't have to sift through extra information.'' 
The feedback shows the complementary strengths of both in supporting varying needs.
\section{Ablation Study}

In addition to comparing \name user interface with the baseline approach, we conducted a detailed analysis of the key features in \name to evaluate their usefulness. To achieve this, we performed an ablation study, evaluating two essential features: ranking and linking. Specifically, we removed one of the features to create an ablated version of \name, allowing evaluation of its usefulness and accuracy in comparison to the full version of \name.

\subsection{Experimental Design}

We evaluate \name with two ablated versions: (a) \name w/o linking removes the links in the middle, and (b) \name w/o ranking removes the color encoding on claims.
We further recruited 36 human participants on Prolific (18 females, 1 preferred not to say gender; average age is 31.9, $SD=6.6$; all native speakers), using three tools with different tasks in a counterbalanced order. Each participant was asked to do three tasks using each tool, resulting in a total of nine comparisons. At the end of the study, participants rated the usefulness of each tool and left comments based on their usage experience. Each of the participants received £6 for the task.

\subsection{Results}

We analyze the user ratings, feedback accuracy, and participants' comments to understand the effectiveness of the key features.

\begin{figure}[!t]
\centering
  \includegraphics[width=\linewidth]{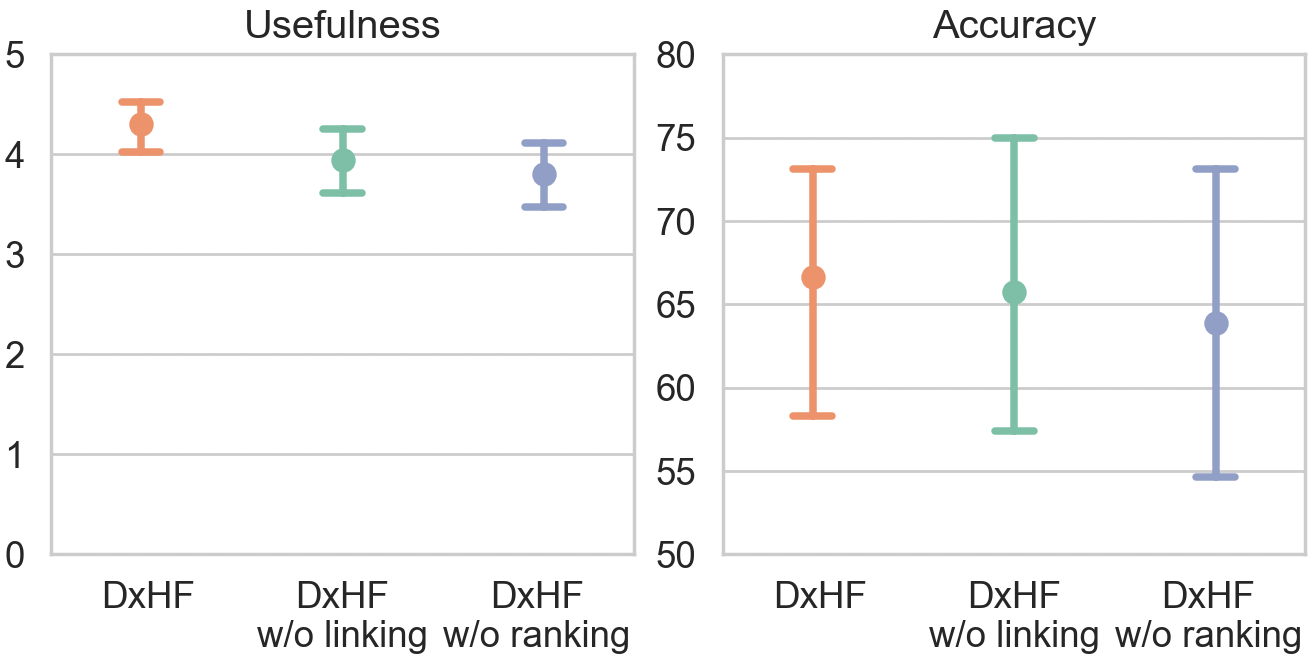}
  \caption{The ablation study results illustrate that \name with full features is the most useful tool and achieves the highest accuracy.
  \rv{The initial accuracy is 50\%, indicating that the preferences were estimated by random guessing.}
  The error bars represent 95\% confidence interval.
  }
  \label{fig:ablation}
\end{figure}

\paragraph{\name performs better than all ablated versions}
Figure~\ref{fig:ablation} illustrates the participants' ratings for usefulness and the accuracies of their feedback.
\name achieves the highest rating in usefulness ($M=4.31, SD=0.79$). \name w/o linking is the second ($M=3.94, SD=0.98$) and \name w/o ranking is the third ($M=3.81, SD=1.01$). 
A repeated measures ANOVA indicates that there is a statistically significant effect of condition on usefulness ($F(2,70) = 4.0093, p = 0.0225$).
Regarding the accuracy of the feedback, the order of the three methods is the same (\name 66.7\%; \name w/o linking 65.7\%; \name w/o ranking 63.\%). 
% All methods outperform the baseline.
% reported in Section~\ref{sec:userstudy}.
% Although no significant differences in the accuracy scores, all of the accuracies are higher than the baseline in Section~\ref{sec:userstudy}.

\paragraph{Partcipant Feedback}
Participants expressed appreciation widely for \name's use of color encoding and linking. 
The combination of these features enabled participants to quickly identify important information and understand logical connections across different responses.
Specifically, the color coding was beneficial for locating key information, with one participant noting,  ``The font colours helped me to understand and point out certain information more easily''.
The linking feature was especially valued for facilitating direct, topic-specific comparisons, as another participant stated, ``Links help pick out the parts in the response for that particular topic''. 
When used together, these features created a visually intuitive structure that enhanced both navigation and comprehension. One participant commented, ``I think this tool is highly useful as it combines both visual cues (color font) and navigational aids (links) to help identify and connect relevant claims. It makes it easier to compare different responses and identify key points quickly. It helps in making more informed decisions.''. 
Overall, participants found the integrated highlighting and linking features to be highly effective for improving comparisons.
\section{Discussion}

Our study results emphasize that although advancements in machine learning algorithms are crucial for improving human-AI alignment, progress can also be made by applying principles of HCI. 
The rise of human feedback techniques~\cite{rafailov2023direct, ouyang2022training} has significantly influenced how LLMs are aligned with humans. 
Nevertheless, the effectiveness of these methods critically depends on the quality of human-generated preference data. 
As demonstrated by our findings, the user interface design used for collecting human feedback plays an essential role in this process.
If annotators provide preferences for textual responses they do not fully read or understand, this can inadvertently introduce biases or inaccuracies into the alignment of LLMs~\cite{chen-etal-2024-humans}.
By applying interactive decomposition, our novel user interface \name greatly improves the annotators' capacity to provide high-quality comparative feedback. This improvement was particularly noticeable among annotators facing uncertainty, a situation frequently encountered in real-world annotation tasks.

% However, the benefits observed from interactive decomposition come with some drawbacks.
% such as the risk of fragmenting the annotator's understanding.
% In our crowdsourcing user study, one participant insightfully noted that while DxHF facilitated deeper analysis of individual claims, it also carried some issues for understanding: ``\name allowed me to do a lot more in-depth analysis of the two responses against each other. However, the danger with \name is that you do not read the text as a whole to get the overall impression of what is being said. Analysis on its own can be misleading without also reading the entire text separately.'' 
% To address this issue, future research should explore a broader design space for interactive decomposition beyond what we investigated in this study, such as the alignment visualization design space~\cite{yousef2020survey}.
% Additionally, future research could also explore decompositions emphasizing different alignment dimensions, such as helpfulness, honesty, and harmlessness (HHH)~\cite{bai2022training}.

\rv{
We have observed that using \name may alter the alignment outcomes, which presents both advantages and disadvantages. On the positive side, it can help reduce the impact of noisy or rushed judgments that can negatively affect alignment. Our study found that it decreases variability in responses, meaning that users' outputs are more aligned with the consensus view.
However, there are some drawbacks to consider. A key concern is whether \name might introduce its own bias. If that is the case, it would be attributable to the LLM that is used to pick the highlighted points. We considered this issue during the design process. So we intentionally reduce the divergence in the reading experience by ensuring that all decomposed claims remain semantically faithful to the original text and preserve its narrative order. Additionally, we designed \name to allow full transparency. Users can always access the complete responses in their original format and decide whether to turn on the enhancement features (see Figure~\ref{fig:metaphor}).
The transformation of the text presentation also increases the risk of fragmenting the annotator's understanding.
One participant from the crowdsourcing user study insightfully commented: ``[\name] allowed me to do a lot more in-depth analysis of the two responses against each other. However, the danger with [\name] is that you do not read the text as a whole to get the overall impression of what is being said. Analysis on its own can be misleading without also reading the entire text separately.'' 
}
% It also increases the risk of fragmenting the annotator's understanding.
To address this issue, future research should explore a broader design space for interactive decomposition beyond what we investigated in this study, such as the alignment visualization design space~\cite{yousef2020survey}.
Additionally, future research could also explore decompositions emphasizing different alignment dimensions, such as helpfulness, honesty, and harmlessness (HHH)~\cite{bai2022training}.

Regarding future work, beyond LLM alignment, the interactive decomposition technique introduced by \name could be applied broadly in human-AI interaction to decompose complex human tasks into simpler ones. As AI systems grow in complexity, there is an increasing demand for effective interfaces that facilitate transparent, meaningful human oversight.
\rv{
The potential impact of DxHF opens the door to fine-tuning cases where expert opinions are necessary, particularly for information-intensive reading tasks. With the redesign, it could support comparisons between human-written text and AI-generated content, illustrating how AI revises human-written text. Additionally, it could enable comparisons between articles written by humans, such as two news pieces from different sources.
}
The paper presents a promising avenue for HCI researchers to contribute more directly to ongoing AI research to enhance human-AI interaction.

% Limitation: incorrect annotations
\section{Conclusion}

In this work, we study the decomposition principle to improve the quality of human feedback for LLM alignment. We build and open-source an intelligent user interface \name based on the decomposition principle.
Evaluations conducted through a technical evaluation, a crowdsourcing user study, and an ablation study show that breaking down complex texts into clear, individual claims significantly improves the accuracy of human feedback, particularly for annotators who may experience uncertainty during comparison tasks. 
The finding highlights the potential of decomposition as a robust strategy for improving human-AI alignment. 
The study presents a promising opportunity for HCI researchers to contribute more directly to AI research.

% Based on this principle, we introduced \name, an interactive user interface designed to be easily integrated into existing AI-based chat applications to facilitate human feedback. 
% Evaluations conducted through a technical evaluation, a crowdsourced user study, and an ablation study show that decomposition leads to increased accuracy across various annotators.

\begin{acks}
DS and AO were supported by the Research Council of Finland (FCAI: 328400, 345604, 341763; Subjective Functions 357578) and the ERC (AdG project Artificial User: 101141916.).
TW was supported through a grant from the Swedish e-Science Research Centre (SeRC).
\end{acks}

\bibliographystyle{ACM-Reference-Format}
\bibliography{reference}

\clearpage
\appendix
\section{Prompts}
\label{sec:prompt}

\subsection{Decomposition Prompt}
\label{sec:decompose_prompt}
We break down the sentence into individual claims automatically, by feeding the sentence to GPT4 (gpt-4o-2024-08-06). The full few-shot prompt, highly referred to the prompts used by \cite{min2023factscore} and ~\cite{gu2024ai}, is designed as follows:

\begin{quote}
    ``\textit{Please break down the given sentence into independent claims. Extract all the claims from a given sentence by copying the words from the original text. A sentence may contain multiple claims. 
    Each claim should try to be of the form <subject> <predicate> <object>, and should have the first occurrence of any pronouns replaced by their antecedents.
    Each claim is a sentence composed of the words from the original sentence. It should not change the meaning in the sentence. It should only copy text, not add new words.
    }
\end{quote}

\begin{quote}
    \textit{Please break down the given sentence into independent claims: You can then add water and mix everything until you have a firm dough}
\end{quote}

\begin{quote}
    \textit{Claim: You can then add water}
\end{quote}

\begin{quote}
    \textit{Claim:  You can mix everything until you have a firm dough}
\end{quote}

\begin{quote}
    \textit{Please break down the given sentence into independent claims: That’s why the driver needs to be paying attention, and must still be able to see clearly, and still must have his or her own ideas about what to do.}
\end{quote}

\begin{quote}
    \textit{Claim: driver needs to be paying attention}
\end{quote}

\begin{quote}
    \textit{Claim: driver must be able to see clearly}
\end{quote}

\begin{quote}
    \textit{Claim: driver must have his or her own ideas about what to do}
\end{quote}

\begin{quote}
    \textit{Claim: driver must have ideas about what to do}
\end{quote}

\begin{quote}
    \textit{\{few more shot instructions if needed\}}
\end{quote}

\begin{quote}
    \textit{Please break down the given sentence into independent claims: \{sentence\}}
    ''
\end{quote}

\subsection{Keyword Prompt}
\label{sec:keyword_prompt}

We summarize the keyword for two claims down automatically by GPT4 (gpt-4o-2024-08-06). The prompt is a zero-shot design.

\begin{quote}
    ``
    \textit{Given the conversation as the context: \{Query\}.}
\end{quote}

\begin{quote}
    \textit{Here are two claims: \{Claim 1\} and \{Claim 2\}.}
\end{quote}

\begin{quote}
    \textit{Please summarize the two claims. No other words.}
    ''
\end{quote}

\subsection{Synthetic User Prompt}
\label{sec:synthetic_prompt}

In the simulation study, we input the conversation as a query and a sentence to GPT-4 (gpt-4o-2024-08-06) to score the helpfulness of the sentence in relation to the query.

\begin{quote}
    ``\textit{Given a query and a sentence. Your duty is to score the helpfulness of the sentence regarding the conversation.}
\end{quote}

\begin{quote}
    \textit{Here is the conversation \{Query\}.}
\end{quote}

\begin{quote}
    \textit{Please score the sentence: \{Claim\}.}
\end{quote}

\begin{quote}
    \textit{The score is a float number from 0 to 1, where 1 means extremely helpful, and 0 means not helpful at all. Please return the score. Ensure you only return a score from 0 to 1.}
    ''
\end{quote}

\section{Alternative Designs}
\label{sec:alternative_designs}

\begin{figure}[htbp]
  \centering
  \begin{subfigure}{0.42\linewidth}
    \centering
    \includegraphics[width=\linewidth]{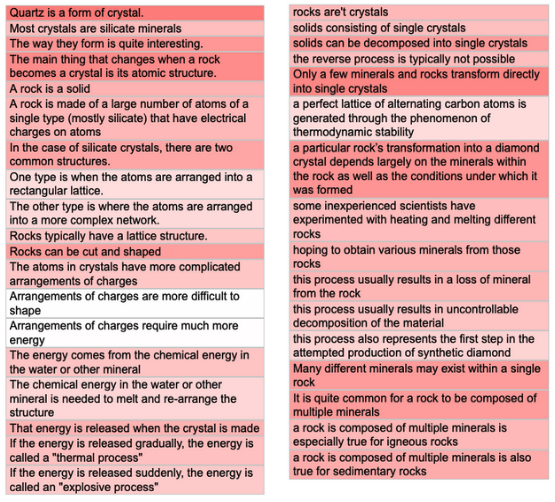}
    \caption{Background color}
  \end{subfigure}
  \hfill
  \begin{subfigure}{0.56\linewidth}
    \centering
    \includegraphics[width=\linewidth]{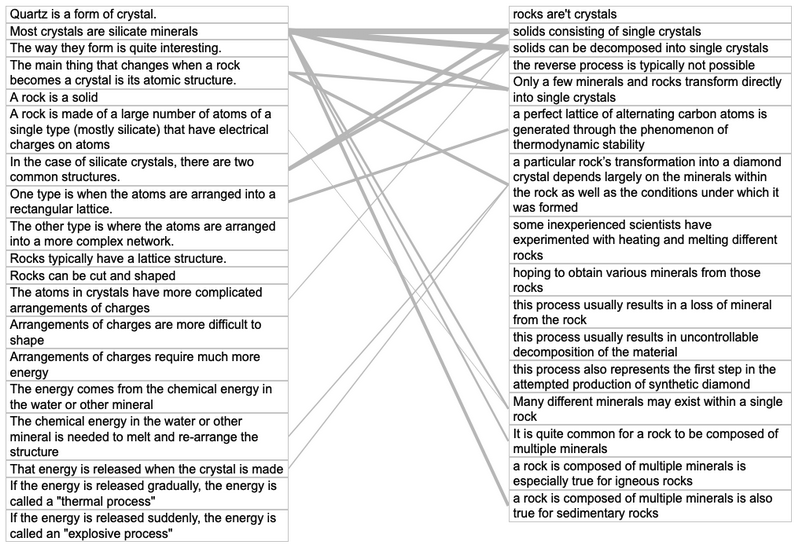}
    \caption{Direct links}
  \end{subfigure}
  
  \caption{The alternative designs for the claim annotations}
  \label{fig:alternative_designs}
\end{figure}

We tested design alternatives, such as using background colors to indicate relevance (Figure~\ref{fig:alternative_designs}-a) or utilizing direct links to illustrate connections (Figure~\ref{fig:alternative_designs}-b). However, both design options led to increased distraction due to visual clutter.

\section{Study Interface}
\label{sec:study_interface}

We present the study interface that crowdworkers use to interact with \name. They receive an introduction and can explore the user interface, after which they provide their feedback along with the certainty of their responses.

\begin{figure}[!h]
\centering
  \includegraphics[width=\linewidth]{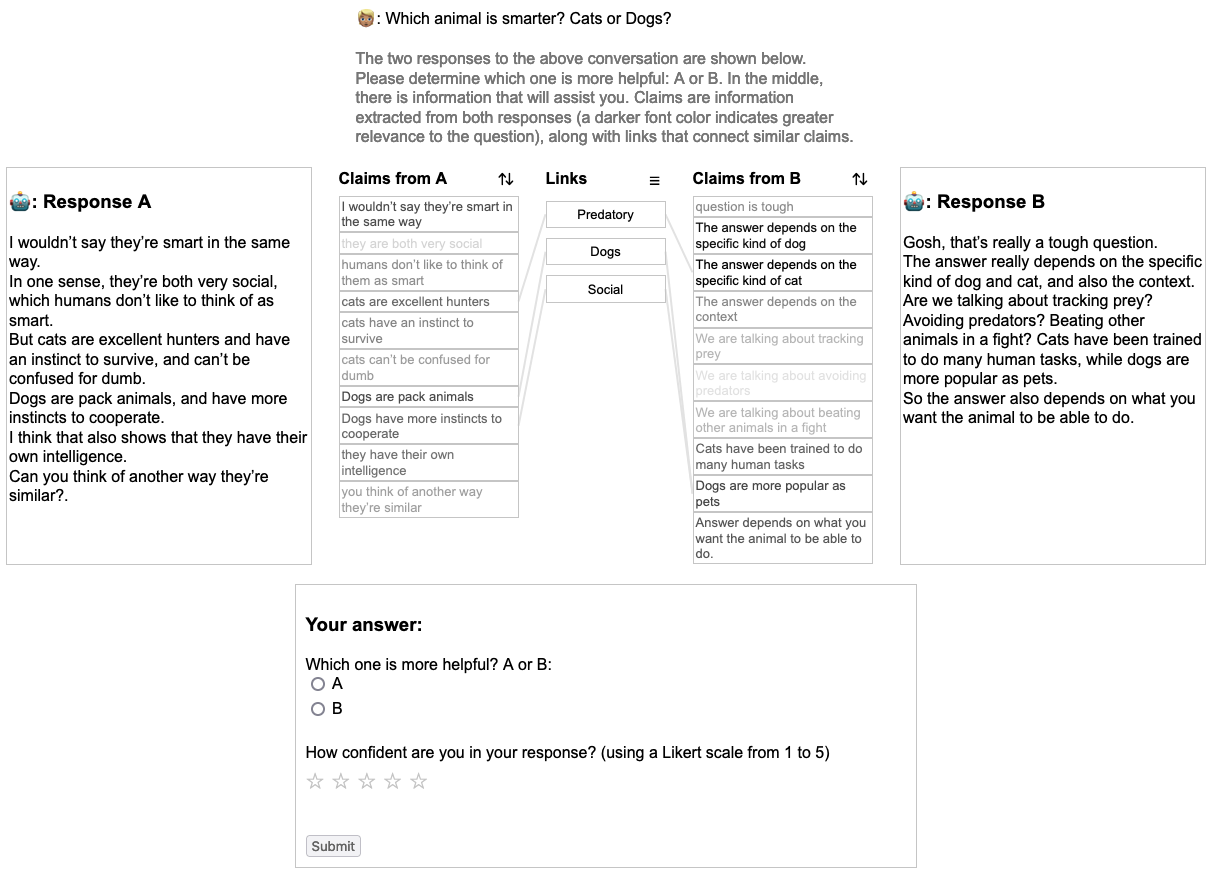}
  \caption{The screenshot of the user interface used in the empirical study for the annotators on Prolific.
  }
  \label{fig:screenshot}
\end{figure}

\end{document}